\documentclass[sigconf]{acmart}
\usepackage{subcaption}
\usepackage{enumitem}
\usepackage{longtable}
\usepackage{makecell}
\usepackage{float}

\usepackage{array}
\usepackage{amsmath}
\usepackage{graphicx}
\usepackage{microtype}

\AtBeginDocument{%
  }

\setcopyright{acmlicensed}
\copyrightyear{2025}
\acmYear{2025}
\acmConference[FAccT '25]{The 2025 ACM Conference on Fairness, Accountability, and Transparency}{June 23--26, 2025}{Athens, Greece}
\acmBooktitle{The 2025 ACM Conference on Fairness, Accountability, and Transparency (FAccT '25), June 23--26, 2025, Athens, Greece}\acmDOI{10.1145/3715275.3732045}
\acmISBN{979-8-4007-1482-5/2025/06}

\begin{document}
\title[Measuring Perceptions and Behavioral Effects of Minoritized Anthropomorphic Cues in LLMs]{Not Like Us, Hunty: Measuring Perceptions and Behavioral Effects of Minoritized Anthropomorphic Cues in LLMs}


\author{Jeffrey Basoah}
\authornote{Both authors contributed equally to this research.}
\email{jeffkb28@uw.edu}
\affiliation{%
  \institution{University of Washington}
  \city{Seattle}
  \state{Washington}
  \country{USA}}

\author{Daniel Chechelnitsky}
\authornotemark[1]
\email{dchechel@andrew.cmu.edu}
\affiliation{%
  \institution{Carnegie Mellon University}
  \city{Pittsburgh}
  \state{Pennsylvania}
  \country{USA}
}

\author{Tao Long}
\email{long@cs.columbia.edu}
\affiliation{%
  \institution{Columbia University}
  \city{New York}
  \state{New York}
  \country{USA}
}

\author{Katharina Reinecke}
\email{reinecke@cs.washington.edu}
\affiliation{%
  \institution{University of Washington}
  \city{Seattle}
  \state{Washington}
  \country{USA}
}

\author{Chrysoula Zerva}
\email{chrysoula.zerva@tecnico.ulisboa.pt}
\affiliation{%
  \institution{Instituto de Telecomunicações, \\IST 
  - University of Lisbon}
  \city{Lisbon}
  \country{Portugal}
}

\author{Kaitlyn Zhou}
\email{katezhou@stanford.edu}
\affiliation{%
  \institution{Stanford University}
  \city{Stanford}
  \state{California}
  \country{USA}
}

\author{Mark Díaz}
\email{markdiaz@google.com}
\affiliation{%
  \institution{Google Research}
  \city{Mountain View}
  \state{California}
  \country{USA}
}

\author{Maarten Sap}
\email{msap2@andrew.cmu.edu}
\affiliation{%
  \institution{Carnegie Mellon University}
  \city{Pittsburgh}
  \state{Pennsylvania}
  \country{USA}
}

\renewcommand{\shortauthors}{Basoah et al.}

\begin{abstract}
As large language models (LLMs) increasingly adapt and personalize to diverse sets of users, there is an increased risk of systems appropriating \textit{sociolects}, i.e., language styles or dialects that are associated with specific minoritized lived experiences (e.g., African American English, Queer slang). 
In this work, we examine whether sociolect usage by a LLM agent affects user reliance on its outputs and user perception (satisfaction, frustration, trust, and social presence). 
We designed and conducted user studies where 498 African American English (AAE) speakers and 487 Queer slang speakers performed a set of question-answering tasks with LLM-based suggestions in either standard American English (SAE) or their self-identified sociolect.
Our findings showed that sociolect usage by LLMs influenced both reliance and perceptions, though in some surprising ways.
Results suggest that both AAE and Queer slang speakers relied more on the SAELM, and had more positive perceptions of the SAELM. 
Yet, only Queer slang speakers felt more social presence from the QSLM over the SAE one, whereas only AAE speakers preferred and trusted the SAELM over the AAE one.
These findings emphasize the need to test for behavioral outcomes rather than simply assume that personalization would lead to a better and safer reliance outcome.
They also highlight the nuanced dynamics of minoritized language in machine interactions, underscoring the need for LLMs to be carefully designed to respect cultural and linguistic boundaries while fostering genuine user engagement and trust.
 
\end{abstract}





\maketitle

\section{Introduction}


With large language models (LLMs) being applied in many domains \cite{milanez2023impact, lee2021application, ghimire2024generative, hancock2020ai, arora2024psychological} 
and interacting with more diverse users from different backgrounds \cite{pandey-etal-2025-culturally}, there have been increased calls for them to adapt better to users and make them more human-like \cite{wang2024understanding}. For example, LLMs are increasingly being personalized to users' language style \cite{salemi2024lamp} and being designed to communicate in more polite ways \cite{ou2024dialogbench}.
However, while this increased adaptation and human-likeness can improve user experience \cite{simas2024human}, this causes various risks such as anthropomorphization through personalization and over-reliance \cite{kirk2023personalisationboundsrisktaxonomy,zhou2024relaiinteractioncenteredapproachmeasuring}.



A significant challenge is that naively adapting LLMs to all users can significantly backfire, particularly in the context of users who are from a different culture or from a minority background.
Many minority users speak \textit{sociolects}, i.e., culturally specific dialects or speech forms \cite{wolfram2004varieties}, such as African American English (AAE) or Queer slang; their language style is deeply intertwined with their lived experience \cite{green2002african, simes2005gay}.
For humans to naively adopt the speaking style and mannerisms of a minority interlocutor can be seen as cultural appropriation and offensive \cite{matthes2019cultural}, raising the question of whether anthropomorphization of LLMs by making them personalize and mimic sociolects of minority users (i.e., making LLMs use anthropomorphic cues) poses similar risks \cite{nigatu2024searched, basoah2025aimimicpeopleunderstanding}.

In this work, we empirically explore this question by analyzing perceptions of sociolect adaptation in LLMs and examining the behavioral impacts these adaptations have on users, adding to the growing body of FAccT and adjacent work that highlights the negative consequences of overlooking individual identity and cultural understanding in technology design \cite{ hada2024akal, bella2024tackling}.
We specifically examine how LLM agents that use minoritized 
\textbf{anthropomorphic cues (i.e., sociolects)} impact the user experience, investigating:
\begin{itemize}
    \item \textbf{RQ1}: Whether a user's \textit{reliance} on the agent depends on whether the LLM agent speaks the user's sociolect or not, to examine possible risks of overreliance due to anthropomorphization \cite{zhou2024relaiinteractioncenteredapproachmeasuring}.
    \item \textbf{RQ2}: Whether a user's \textit{perception} varies depending on whether the agent uses sociolect or not, to characterize possible risks of cultural appropriation \cite{matthes2019cultural}.
    \item \textbf{RQ3}: Whether \textit{perception} factors (e.g., trust, frustration) are associated with user \textit{reliance}, to explore variation and association between reliance and other factors.
\end{itemize}

\looseness=-1
To explore these questions, we devised a within-subjects user study where participants answered factual questions about two videos with the help of suggestions given either by a sociolect-using LLM agent or by a standard American English (SAE) LLM agent, inspired by setups from HCI and natural language processing (NLP) \cite{bansal2019beyond,zhou2024relaiinteractioncenteredapproachmeasuring}.
We assessed users' reliance on each agent behaviorally and evaluated users' perceptions across four key dimensions: trust, satisfaction, frustration, and social presence along with a pairwise judgment and rationale of which agent they preferred.
We focused our investigations on AAE and Queer slang, motivated by the history of negative experiences Black and LGBTQ+ users with artificial intelligence (AI) systems \cite{dodge-etal-2021-documenting, harris2022aae, ovalle2023trans}, as well as the authors' own positionality.



Our results from 498 AAE speakers and 487 Queer slang speakers show that sociolect usage by LLMs influences both reliance and user perception in most cases, highlighting the nuanced relationship between language, behavior, and user perceptions in machine interactions. 
Notably, AAE participants preferred and relied more on the SAE-using LLM, while, surprisingly, Queer slang speakers showed no significant preference. AAE participants trusted, were more satisfied with, and experienced less frustration with the SAE-using LLM. In addition, while Queer slang speakers reported less frustration with the SAE-using LLM, they felt greater social presence with the Queer slang-using LLM. 
Thematic coding of the rationales highlights how participant comments mirrored key trends from the quantitative findings, while offering nuanced insights into user perceptions of sociolect-using LLMs.
We conclude by discussing the implications and risks of sociolect usage by LLMs, emphasizing the need to examine both user perceptions and behaviors. Our findings reveal that users prefer SAE outputs, yet sociolect-using LLMs may be perceived as warmer and friendlier in some cases, underscoring the importance of understanding the historical and social nuances of each sociolect. Furthermore, our work addresses broader issues of linguistic bias and inclusivity, highlighting the challenges of sociolectal adaptation and developing inclusive LLMs that respect and do not appropriate minority cultures.

\section{Background}
To contextualize our investigations into the effect of sociolect usage by LLMs, we first provide background on AAE and Queer slang sociolects, then provide an overview of related work examining minority experiences with LLMs.

\subsection{Sociolects and Biases in LLMs}

A sociolect is a dialect, or variant, that differs from the standard form of a language that is primarily associated with a specific social group~\cite{kiesling2019language, lewandowski2008language}. 
Sociolects can reflect an intersection of several different identities, such as ‘Valley Girl’ spoken predominantly by women of a certain age range and geographical location \cite{leigh2021style}. Within NLP, studies that look at sociolect are often also grouped into dialect or multi-language studies, and there are many applications within socially-aware and low-resource language NLP contexts \cite{ou2024dialogbench, ziems-etal-2023-multi}.

In our study, we examined two specific sociolects: AAE and Queer slang, chosen because of the history of biases against their speakers in NLP, their interconnectedness as sociolects \cite{sap2019risk, deas2023evaluation}, as well as the positionality of many of the authors who identify as speakers of either AAE, Queer slang, or both.

AAE is a widely studied sociolect of American English spoken predominately, but not exclusively, by Black people in the United States (U.S.) \cite{green2002african, grieser2022black, baker2020linguistic}.\footnote{Although some refer to the language variety as African American (Vernacular) English (AAVE) or African American Language (AAL), we opt for the African American English (AAE) terminology based on previous work \cite{dacon2022towards, sap-etal-2022-annotators}.}
AAE has its own distinct grammatical and phonological features that distinguish it from SAE. Despite existing literature on AAE, it is still an underrepresented both in society as well as online spaces, which is associated with the marginalization of Black AAE speakers in predominantly White spaces \cite{alim2016}. 
Biases against AAE have been widely documented in NLP systems \cite{sap2019risk, shearer2019racial} and specifically LLMs \cite{dodge-etal-2021-documenting, ziems-etal-2022-value, xu-etal-2021-detoxifying, harris2022aae}. For instance, a recent study by ~\citet{deas2023evaluation} 
found that current LLMs have difficulty both generating and interpreting AAE due to the lack of Black American representation in LLMs. 
The authors emphasized the potential harm to the AAE-speaking community and called for further development of LLMs that can effectively interact with and understand AAE speakers
~\cite{deas2023evaluation}.

Queer slang, unlike AAE, is a relatively less well-documented sociolect which can be defined by words, phrases, or metaphors predominantly used by LGBTQ+ individuals in the U.S. (e.g., in queer or drag spaces) \cite{simes2005gay,mann2011drag}.\footnote{We use the term Queer slang instead of Gay slang because it encompasses more identities within the LGBTQ+ community. The emergent meaning of the word Queer to be all encompassing of LGBTQ+ individuals can be seen in recent work \cite{worthen2023queer}.}
It is closer to SAE, yet could be defined as a minority sociolect due to the its ties with Queer individuals who speak it \cite{casey2019discrimination}. 
Importantly, Queer slang is heavily interconnected with AAE, as Drag and Ballroom culture as well as prominent Black Queer and femme folks (e.g., RuPaul, T.S. Madison) have historically influenced mainstream Queer culture \cite{yass, krudysz2023translating,laing2021said}.
Recent NLP work has shown documented biases in LLMs against Queer identities and Queer slang 
\cite{dodge-etal-2021-documenting, dhingra2023queer}.
Similar to AAE, little research explores how systems interpret, generate, and classify Queer slang differently from SAE \cite{felkner2023winoqueer}.



\vspace{-3px}
\subsection{Minority Experiences with LLMs}

Language models have shown limitations in serving diverse user groups \cite{bella2024tackling}. Previous studies have highlighted the challenges posed by inherent biases within these systems' training data and processes \cite{kapania2023hunt, muzanenhamo2024chatgpt} --- LLMs often exhibit undesirable behaviors related to ethical issues, particularly towards minorities based on gender and race, leading to suboptimal user experiences, confusion, and technology abandonment \cite{luna2024evaluating, wan2023kelly, Tweetorial_Hook}. 
Additionally, opinion minority groups (e.g., climate change deniers, the alt-right) and educational minority groups tend to have poor user experiences with LLMs \cite{chen2024conversational}. These challenges sparked new regulations and increased public awareness about the associated risks.

Along with linguistic, cultural, gender, and ability biases \cite{bella2024tackling, gadiraju2023disability, lee2024large} many studies have noted that AI systems and LLMs frequently fail to account for Black and LGBTQ+ identities or challenges \cite{dhingra2023queer, ovalle2023trans, mei2023bias, kleiner2024unmasking, hofmann2024ai}.  \citet{mengesha2021culturally}  highlights significant performance disparities in AI systems' understanding of African American language (AAL) are due to their lack of design for the Black experience. Additionally, \citet{ma2024evaluating} shared LLMs tend to generate overly generic or insensitive suggestions, such as advising LGBTQ+ users to come out to unsupportive parents, which risks offending and marginalizing these communities.

\section{Approach: Research Questions \& Variables Studied}

In order to observe how sociolect use of LLMs can affect user's conscious and subconscious perceptions of LLMs, we  explored both behavioral and perceptual effects based on previous work. The rationale for each variable will be discussed in each respective section below.



\subsection{User Reliance (RQ1)}

Our first research question asks: \textbf{Does a user's reliance on the agent depend on whether the LLM agent speaks the user's sociolect or not?}
\textit{Reliance}, as commonly defined, is the degree to which users use the help of the target LLM in their daily lives. This is a useful metric because it provides insight into how much the user implicitly trusts the LLM outputs on a wider scale \cite{zhou2024relying}. We distinguish between self-reported reliance, which would capture participants’ perceptions of how much they relied on the LLM, and behavioral reliance, which is based on observable actions—specifically, whether participants accepted the LLM’s suggestion during the task. While self-reported reliance reflects subjective impressions, behavioral reliance provides a more objective measure of actual trust-based decision-making. We prioritize behavioral reliance in our analysis as it more directly reflects participants' willingness to defer to the LLM in practice, beyond just what they say. Existing works that measure reliance have also observed that reliance more closely measures how a user depends on, understands, and would utilize the system in the future \cite{seo2024chacha, ma2024evaluating}. Systems that express warmth in their outputs are correlated with higher reliance \cite{zhou2024relaiinteractioncenteredapproachmeasuring} and also tend to be generally preferred \cite{gilad2021warmth}. We propose that the use of a user's sociolect will garner greater user reliance in that LLM (\textbf{H1}).

\subsection{User Perception and Pairwise Preference (RQ2)}

Our second research question asks: \textbf{Does a user's perception and preference vary depending on whether the agent uses sociolect or not?} 
We specifically consider four perception dimensions for each agent, as well as a pairwise preference judgment, outlined below.


We measure user \textit{trust}, defined as the willingness to be vulnerable based on the expectation of beneficial actions from another party~\cite{HARRISONMCKNIGHT2002297, socialdynamics, gefen2003trust}. We hypothesize that the use of sociolects by LLMs enhances user trust (\textbf{H2a}), supported by works showing anthropomorphic cues boost trust in AI systems, including conversational agents and LLMs \cite{cohn2024believing, wu2024sunnie, deshpande2023anthropomorphization, banovic2023being, jiang2023make}.


User \textit{satisfaction} evaluates session experience, encompassing satisfaction with both LLM and user performance \cite{hart1988development}. Research suggests that anthropomorphic cues, such as emotional and auditory elements, improve interaction satisfaction \cite{xie2023does, klein2023impact}. We hypothesize that adapting LLMs to user's sociolect will increase satisfaction (\textbf{H2b}).

Our third variable, user \textit{frustration}, is crucial in creativity support research as it tracks cognitive load challenges, encompassing feelings of insecurity, stress, and annoyance~\cite{hart1988development}. AI tools can mitigate frustration by providing assistance, reducing mental load, and supporting complex tasks~\cite{Tweetorial_Hook, 10.1145/3613904.3642868, longitudinalAIinteraction, disco}. We hypothesize that sociolect-equipped LLMs, by mirroring a conversation partner’s language style, lower user frustration and perceived task workload (\textbf{H2c}) \cite{markl2023language, spillner2021talk}.

Our last perception variable, \textit{social presence} refers to a user's perceived sense of connectedness with the system they interact with \cite{schutt2023effect}. We hypothesize that users  feel a stronger social presence with sociolect-using LLMs (\textbf{H2d}), as prior work indicates increased social presence in interactions with AI heightened social and cultural awareness \cite{chen2024conversational, venkit2023nationality}.

For pairwise \textit{preference} judgments and \textit{rationales}, users evaluate which LLM agent they prefer---whether sociolect-using or SAE---and provide reasons in free text for their choice. We hypothesize that users will prefer the agent using their sociolect, similar to previous chatbot research  adopting linguistic alignment ~\cite{spillner2021talk} (\textbf{H2e}).


\subsection{Exploratory Findings (RQ3)}

Lastly, our third research question asks: \textbf{What perception factors (e.g., trust, frustration, explicit preference) are associated with user reliance?}
As an exploratory research question, we hypothesize that there will be associations between perception variables, including explicit user preference and reliance \textbf{(H3)}.

\section{Methods}

To investigate how sociolect-specific language models influence user perception and reliance, we conducted a within-subjects experiment with two groups of participants (AAE speakers, Queer slang speakers;~\S\ref{ssec:recruitment}), in which participants performed two sequential, video-based question answering tasks with the help of a simulated LLM -- either a sociolect-producing or a control SAE LLM (\S\ref{ssec:experimental-procedure}). 





\subsection{Recruitment and Screening Procedure}\label{ssec:recruitment}

We recruited participants using Prolific, an online platform tailored for academic research~\cite{palan2018prolific}. Prolific allowed us to target specific participant groups based on eligibility criteria. Participants had to be at least 18 years old and reside within the U.S., ensuring a shared cultural exposure to the sociolects under investigation. The study description provided to potential participants on Prolific outlined the procedure: participants would watch two short videos (each under one minute) and evaluate various aspects of their experience.

For the African American English LLM (AAELM) setup, participants were required to self-identify as Black American based on Prolific’s demographic question: ``Please indicate your ethnicity (i.e., peoples’ ethnicity describes their feeling of belonging and attachment to a distinct group of a larger population that shares their ancestry, color, language, or religion).'' For the Queer slang LLM (QSLM) setup, participants were required to identify as members of the LGBTQ+ community on Prolific, as indicated by their response to the question: ``Do you identify yourself as part of the LGBTQ+ community?'' We later ensured their familiarity with the respective sociolect as part of the main study (\S\ref{ssec:experimental-procedure}).

To ensure fair compensation and value participants' time, we offered an hourly rate of \$15, which was clearly communicated during recruitment to maintain transparency and encourage participation. The study protocol, including recruitment and screening procedures, was reviewed and received exempt status by {the Carnegie Mellon University Institutional Review Board (IRB). This oversight ensured that all participant interactions adhered to the highest ethical standards, particularly concerning privacy, confidentiality, and participant welfare.

\subsection{Experimental Procedure}\label{ssec:experimental-procedure}

The study employed a within-subjects design for each participant group, as depicted in Figure~\ref{fig:study flow}. Participants began by completing a consent form (see Appendix~\S\ref{ssec:consent-form}). Non-identifiable demographic information, including age, gender identity, and education level, was collected. Participants were also asked about their familiarity with LLMs, including frequency of use, confidence in understanding, and typical use cases (see Appendix~\S\ref{ssec:demographics}).

\begin{figure*}[htbp]
    \centering

    \includegraphics[width=\linewidth]{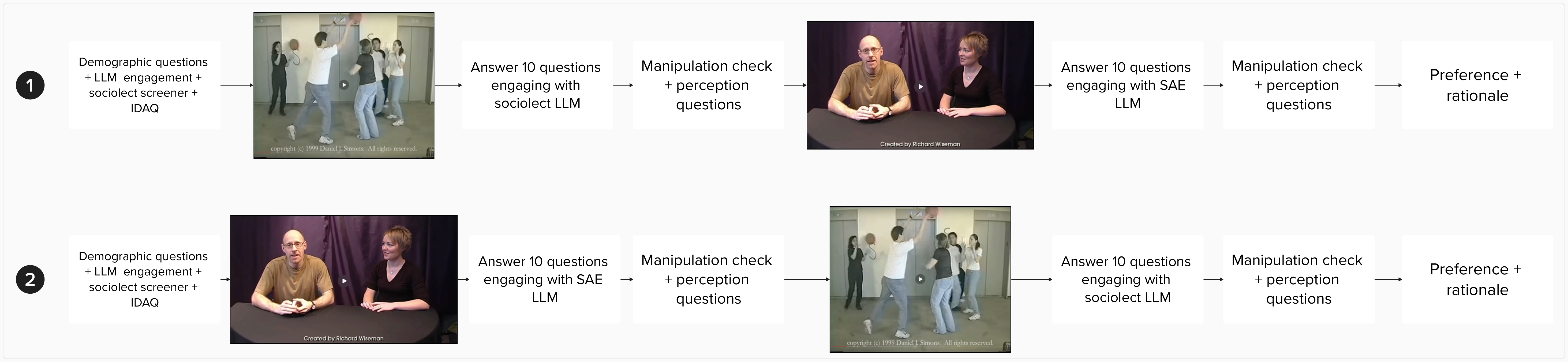}
    \caption{Study flow diagram illustrating the two experimental scenarios. In both scenarios, participants begin by answering demographic questions, completing a sociolect screener, and the IDAQ. In Scenario 1 (top row), participants first engage with a sociolect LLM after watching a basketball-passing video (Daniel Simons' "Gorillas in Our Midst"), followed by a manipulation check, perception questions, and a preference question. In Scenario 2 (bottom row), participants first engage with a SAE LLM after watching a different video (Richard Wiseman’s "Color-Changing Card Trick"), followed by the same sequence of tasks.}
    \label{fig:study flow}
    \Description{Diagram outlining the two study conditions. In Condition 1, participants first interact with a sociolect LLM and then a Standard American English LLM; in Condition 2, the order is reversed. Each condition includes demographic questions, LLM engagement tasks, manipulation checks, perception questions, and a final preference rationale.}
\end{figure*}

To ensure participants were familiar with the sociolect being tested, a sociolect screener was implemented. Participants were screened using a 5-point Likert scale to assess their frequency of sociolect use, confidence in understanding, and likelihood of using example sociolect phrases. Example phrases for the AAELM and QSLM setups were sourced from established corpora and are detailed in the Appendix~\S\ref{ssec:sociolect-screener}.

Participants who passed the screener completed the Individual Differences in Anthropomorphism Questionnaire (IDAQ)~\cite{waytz2010sees} to assess their tendency to anthropomorphize non-human entities (see Appendix~\S\ref{ssec:idaq}). We include measurements of IDAQ to explore relationships between participant reliance and LLM perceptions and their general tendency to anthropomorphize non-human agents. We initially included it to explore relationships between user reliance or LLM perceptions and a user's general tendency to anthropomorphize non-human agents but found no significant results. Results and hypotheses as to why there was no significant association can be found in the Appendix~\S\ref{ssec:resulttables}.

Participants were then shown a snippet from one of two videos designed to test selective attention: Daniel Simons’s ``Gorillas in Our Midst''~\cite{simons1999gorillas}\footnote{\url{https://www.youtube.com/watch?v=vJG698U2Mvo}} or Richard Wiseman’s ``The Colour Changing Card Trick''~\cite{wiseman2007quirkology}.\footnote{\url{https://www.youtube.com/watch?v=v3iPrBrGSJM}} 
Selective attention videos captivate participants by presenting detail-rich scenarios that demand focused observation, ensuring engagement and attentiveness~\cite{simons1999gorillas}. By using videos with highly specific and difficult-to-notice details, participants were placed in a situation in which reliance on the LLM's suggestions became a meaningful decision~\cite{Paas2003,Sweller1988}. This allowed us to evaluate not only the participants' perception of the LLM but also their decision-making when balancing their own knowledge against the LLM's suggestions.

After each video, participants answered questions about specific events while receiving suggestions from their respective LLM agent in SAE or the assigned sociolect. Questions were designed to be difficult, focusing on specific video details (e.g., “What color watch was the woman wearing?”). Question difficulty was pre-assessed by an independent participant pool via Prolific, using a 7-point Likert scale. Only questions rated on average between "very difficult" and "difficult" were selected; an excerpt of a difficulty survey can be found in Appendix~\S\ref{ssec:difficulty-survey}.

Inspired by \citet{zhou2024relying}, during the video assessment, participants could choose to accept the LLM's suggested responses (“Use LLM's Response”) or rely on their own knowledge (“I'll figure it out myself”). There was minimal framing ("You will interact with a LLM agent") to allow participants an opportunity develop mental models based solely on the language the systems used. Participants were told they would earn points based on their ability to correctly reject incorrect LLM suggestions or rely on accurate ones. Although no points were actually calculated, this mechanism encouraged thoughtful decision-making and discouraged blind reliance on the LLM~\cite{bansal2019beyond, zhou2024relying}. The order of LLM assignment (sociolect vs. SAE) and video presentation was counterbalanced to control for order effects (see Figure~\ref{fig:study flow}). 

After completing the video task, participants filled out a follow-up questionnaire evaluating several dimensions of their experience. These included \textit{trust} (7-point Likert scale adapted from~\cite{gefen2003trust}), \textit{satisfaction} (7-point Likert scale adapted from~\cite{wang2024lave}), \textit{frustration} (7-point Likert scale adapted from~\cite{hart1988development}), and perceived \textit{social presence} with the LLM (7-point Likert scale adapted from~\cite{jiang2023make}) (see Appendix~\S\ref{ssec:follow-up} for questionnaire). To assess reliance, we adapted our approach from~\cite{zhou2024relying}, which in turn was based on the method used by~\cite{bansal2019beyond}. We operationalized behavioral reliance as the participant’s willingness to accept the LLM’s suggestion during the task, i.e., the proportion of times they selected "Use LLM's response."

Additionally, a manipulation check assessed whether participants perceived the LLM as accurately using the intended sociolect (see Appendix~\S\ref{ssec:manipulation-check} for further details). The study concluded with a post-study LLM agent \textit{preference} question, where participants were asked to choose between the sociolect or SAELM and provide a \textit{rationale} or explanation for their preference.

\subsection{Templated LLM Suggestions}\label{ssec:templateLLMsuggestions}
We designed three distinct simulated LLM agents to generate responses in SAE and sociolects. The AAELM simulated interactions using AAE, while the QSLM simulated interactions using language and slang prevalent in Queer communities. The SAE LLM (SAELM) served as a baseline or control model, using SAE. Importantly, participants in the experiment did not directly interact with live LLMs but instead engaged with responses pre-generated by LLMs. 

Our goal was to create LLM responses that elicited a sense of warmth and medium confidence from participants to observe the effect of sociolect usage on reliance, because previous research has shown that warmth markers can increase user reliance on medium-confidence responses~\cite{zhou2024relaiinteractioncenteredapproachmeasuring}.

\paragraph{SAE Suggestions}
Following the procedure in \citet{zhou2024relaiinteractioncenteredapproachmeasuring}, we obtained a set of SAE warmth phrases from LLMs by prompting popular LLMs to respond to factual questions with warmth prefixes (e.g., ``I'd be happy to help''). 
We merged these LLM-generated warmth prefixes with the confidence prefixes with reliability percentages between 30\% and 70\% (e.g., ``I would say it's''), drawn from Table 6 of \citet{zhou2024relying}.
We compiled a list of 14 warmth phrases and 14 confidence expressions (see Appendix~\S\ref{ssec:aae-phrases-and-markers}), resulting in 196 unique combinations. 

\paragraph{AAE Translation with In-Context Learning}\label{ssec:incontext}

We translated the 196 unique SAE suggestions into AAE using GPT-4’s in-context learning capabilities. This approach aligns with recent advancements in evaluating sociolect-specific text generation~\cite{cheng2023marked, cain2024prompting, jorgensen2015challenges, amadeus2024bridging}. For in-context learning, we sourced AAE examples from CORAAL~\cite{kendall2023corpus}, supplemented with AAE tweet pairs~\cite{groenwold2020investigating} and SAE-to-AAE translations~\cite{deas2023evaluation, kleiner2024unmasking}. The detailed prompt design and implementation process for the in-context learning methodology are described in Appendix~\S\ref{ssec:in-context}.

After translating the 196 warmth-epistemic combinations into AAE, we used Prolific surveys to verify authenticity, asking participants to identify which phrases aligned most closely with AAE (see Appendix~\S\ref{ssec:aaeverification}). We presented groups of 20 phrases to batches of 5 participants at a time and participants were tasked with identifying which sociolect they felt best aligned with the translated phrases. Only phrases that were deemed to represent AAE by at least 4 out of the 5 verifiers were selected for inclusion as AAELM suggestions. This process was repeated until we achieved a diverse selection of 10 sociolect aligned phrases with unique confidence expressions (see Appendix~\S\ref{ssec:AAEtranslation} for final list of AAE translations and corresponding SAE translations.)

\paragraph{Queer Slang Translation with Persona-Based Prompting}

Since no existing corpora provide modern Queer slang or SAE-to-Queer slang translations, we employed persona prompting with GPT-4 to translate SAE suggestions into Queer slang. Persona prompting, where a LLM is guided to emulate the speaking style of specific individuals, has been demonstrated to effectively replicate distinctive linguistic styles~\cite{benharrak2024,ivey2024realroboticassessingllms}. For this study, we selected personas associated with well-known Queer celebrities—RuPaul, Trixie \& Katya, and T.S. Madison—each known for their use of Queer slang~\cite{yass,glover2016redefining,kameswara2024hearer}. Using this approach, we applied the 14 SAE warmth phrases (see Table~\ref{tab:warmth}) to each persona output. Through persona prompting, we generated 42 unique Queer warmth phrases (see Appendix~\S\ref{ssec:queer-phrases-and-markers}). This resulted in 420 unique combinations of Queer warmth phrases.


Using a similar verification procedure as for AAE (see Appendix~\S\ref{ssec:incontext}), we validated Queer slang translations via sociolect verification surveys on Prolific (see Appendix~\S\ref{ssec:queerverification}). Phrases identified by at least 4 out of 5 participants were included as QSLM suggestions (see Appendix~\S\ref{ssec:queertranslation}). Corresponding SAE translations were paired with the same questions as their sociolect counterparts (see Appendix~\S\ref{ssec:queersetup}).



\subsection{Participant Statistics and Demographics}

Out of 1,007 responses, 985 participants successfully passed the sociolect screener, with 498 in the AAELM setup and 487 in the QSLM setup. We included a manipulation check to determine whether participants recognized that the LLM's outputs were indeed in a specific sociolect. Distinguishing between those who recognized the sociolect and those who did not was critical to our analysis. Our manipulation check assessed whether participants perceived the LLM as using the sociolect successfully. This perception was essential for distinguishing between participants who bought into the sociolect attempt and those who did not. To ensure the validity of our findings, we proceeded with quantitative analysis using only the data from participants who passed the manipulation check—those who clearly perceived the intended distinction between the sociolect-speaking LLM and the SAE using LLM. However, for our qualitative analysis, we included responses from all participants, regardless of whether they passed the manipulation check, in order to capture a broader range of perceptions and preferences that were not necessarily contingent on recognizing successful sociolect usage. Among those who passed the manipulation check, those in the AAELM group ranged from 18 to 75 years of age, with 207 identifying as men, 186 as women, and 3 as non-binary and 3 as other. In the QSLM group, ages ranged from 18 to 72 years of age, with 169 identifying as men, 190 as women, and 47 as other (see Appendix~\S\ref{ssec:demotable}).

\section{Findings}

To better understand the impact of sociolect adaptation in LLM interactions, we examined participant reliance, perception, and preferences through a combination of quantitative and qualitative analyses. 


\subsection{RQ1: Does reliance on the agent depend on whether the agent speaks the user's sociolect or not?}

\begin{figure}[!h]
    \centering
    \begin{subfigure}{0.49\linewidth} 
        \centering
        \caption{Participant reliance for\\AAE and SAE.}
        \includegraphics[width=\linewidth]{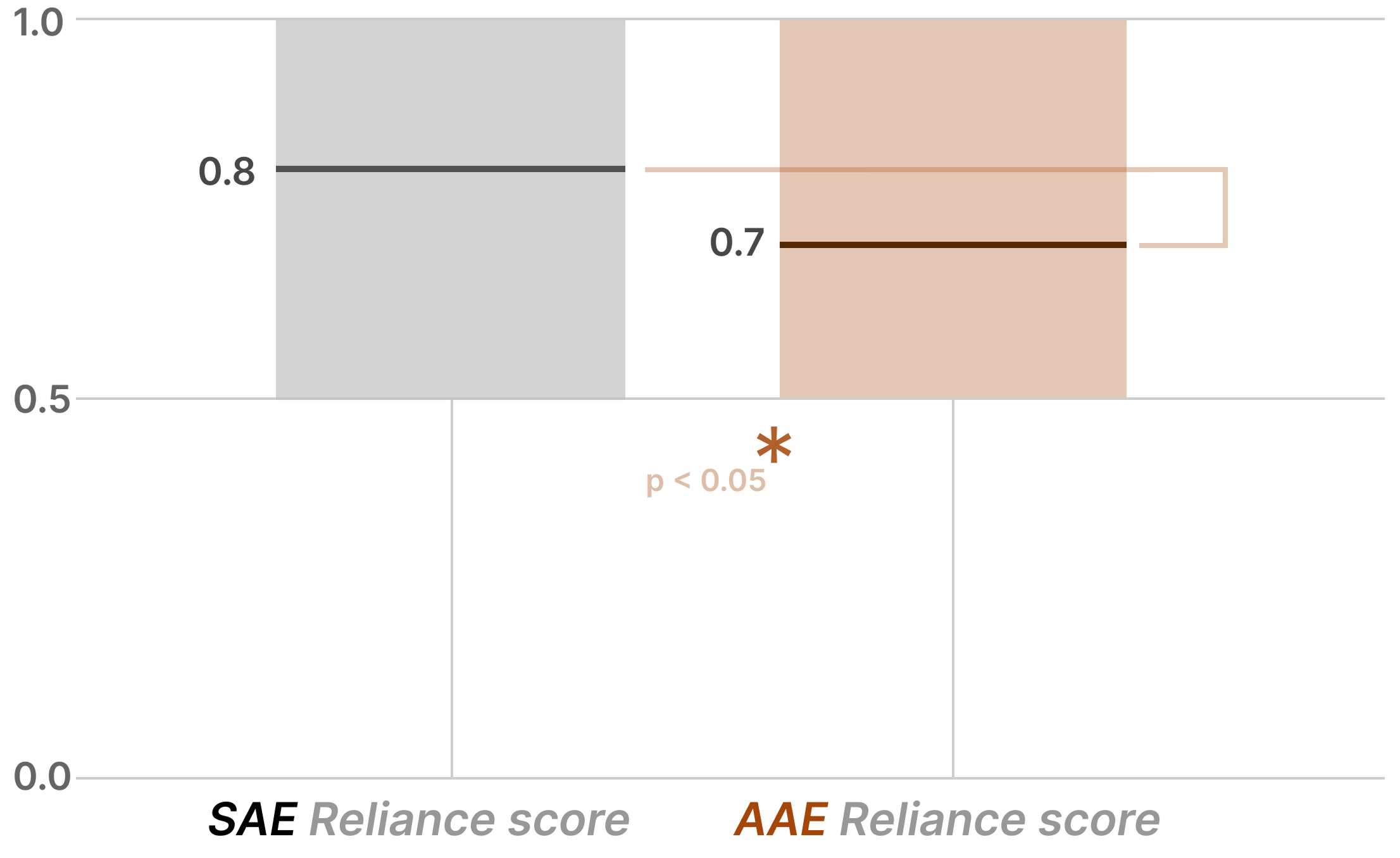}
        \label{fig:aaebox}
    \end{subfigure}
    \hfill
    \begin{subfigure}{0.49\linewidth} 
        \centering
        \caption{Participant reliance for\\Queer slang and SAE.}
        \includegraphics[width=\linewidth]{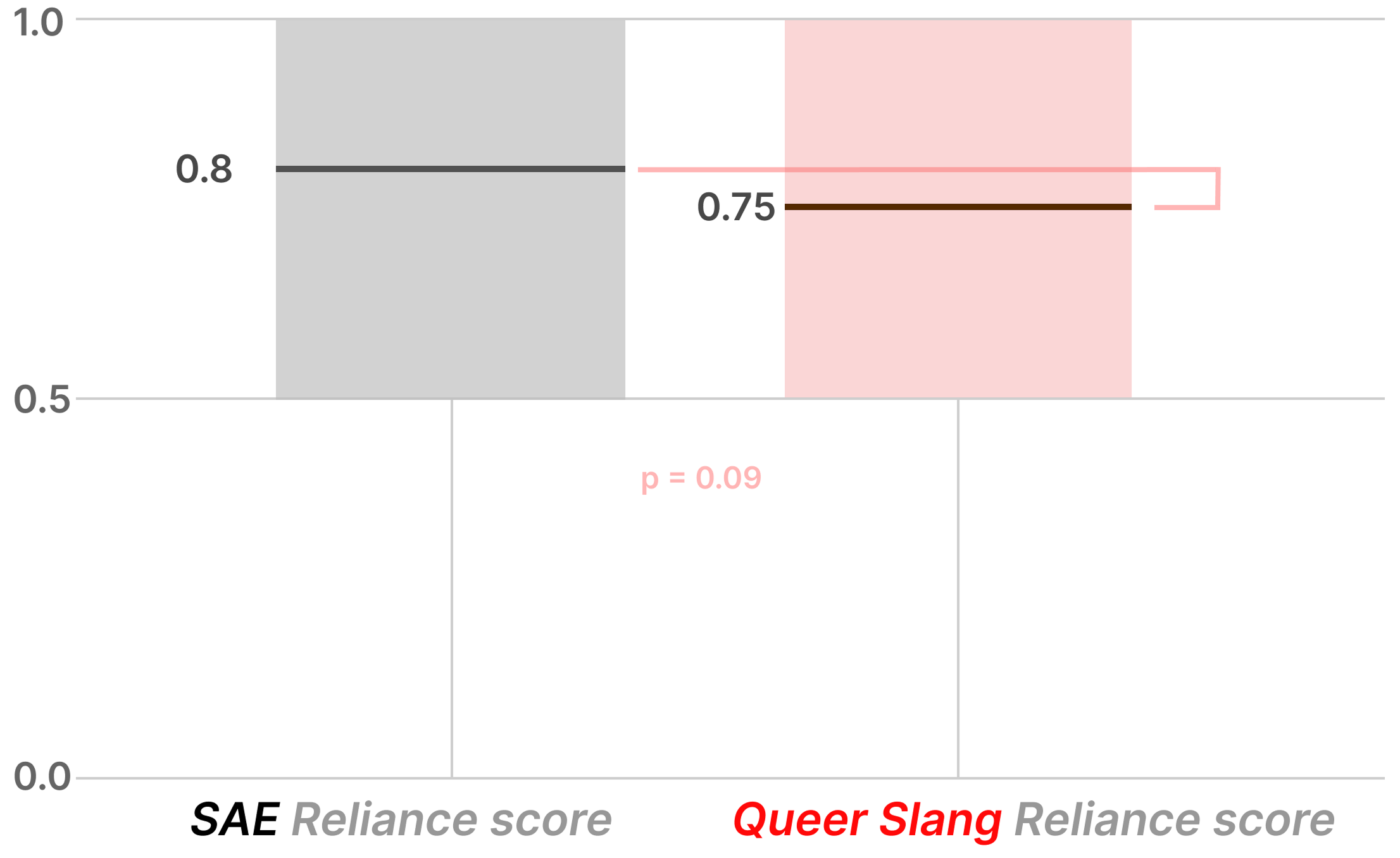}
        \label{fig:queerbox}
    \end{subfigure}
    \caption{RQ1: Participant reliance for different language styles: (a) Boxplot comparing reliance scores for AAE and SAE, showing slightly higher reliance on SAE. (b) Boxplot comparing reliance scores for Queer slang and SAE, with similar ranges and medians for both. Reliance scores range from 0 (low reliance) to 1 (high reliance).}
    \Description{Two bar plots showing participant reliance on standard versus sociolect LLMs. The left plot compares reliance scores between African American English and Standard American English, showing higher reliance on SAE. The right plot compares Queer slang and SAE, showing nearly equal reliance with a slight difference.}
    \label{fig:combined_boxplots}
\end{figure}

To test our first hypothesis (H1: Sociolect usage by LLM leads to higher reliance), we compared reliance using paired t-tests. 


Our analyses showed mixed results. Among AAE-speaking participants, reliance significantly varied depending on whether the agent spoke AAE or SAE ($d=0.12$, $p<0.05$; Fig.~\ref{fig:aaebox}, Table~\ref{tab:rq1reliance}), however, reliance was larger on the SAELM compared to the AAELM, contradicting our hypothesis. For Queer slang speakers, reliance showed a marginal difference depending on whether the agent used Queer slang or SAE ($d=0.08$, $p=0.09$; Fig.~\ref{fig:queerbox}, Table~\ref{tab:rq1reliance}), but this result did not reach statistical significance, and thus, did not support our hypothesis. 



\subsection{RQ2: Does a user’s perception vary depending on whether the agent depend on sociolect usage or not? (across trust, satisfaction, frustration, social perceived presence, and pairwise preference judgment)}

To test our second hypothesis (H2: Sociolect usage by LLM leads to higher trust (H2a), higher user satisfaction (H2b), lower user frustration (H2c), a higher sense of social presence (H2d), and paired preference for the sociolect-using LLM more than the SAELM (H2e), we performed paired t-test comparisons between the relevant variables reported for SAELM and AAELM. 

For AAE speakers, the SAELM consistently was higher than the AAELM across key perception variables. AAE participants trusted the SAELM more than the AAELM ($d=0.19$, $p<0.05$; Table~\ref{tab:rq2}), found it more satisfying ($d=0.2$, $p<0.05$; Table~\ref{tab:rq2}), and experienced less frustration with it ($d=0.21$, $p<0.05$; Table~\ref{tab:rq2}). Interestingly, no significant difference was observed in the perceived social presence between the two agents ($d=-0.05$, $p>0.1$; Table~\ref{tab:rq2}).

Unlike AAE participants, Queer slang participants did not demonstrate a significant difference in trust between the QSLM and the SAELM ($d=-0.03$, $p>0.1$; Table~\ref{tab:rq2}). However, Queer slang participants did perceive significantly greater social presence in the QSLM compared to the SAELM ($d=-0.23$, $p<0.05$; Table~\ref{tab:rq2}). Despite this, satisfaction levels were similar between the two agents ($d=0.07$, $p>0.1$; Table~\ref{tab:rq2}), and participants found the SAELM significantly less frustrating than the QSLM ($d=0.245$, $p<0.05$; Table~\ref{tab:rq2}). 

Examining pairwise perception, we find that AAE participants demonstrated a clear preference for one agent over the other ($d=0.44$, $p<0.05$; Fig.~\ref{fig:aaebar}, Table~\ref{tab:rq2preferredreliance}). In contrast, Queer slang participants showed no significant preference between the two agents ($d=0.009$, $p>0.1$; Fig.~\ref{fig:queerbar}, Table~\ref{tab:rq2preferredreliance}). 

\subsection{RQ3: Are there other factors that influence user reliance?}


To test our third hypothesis (H3: There will be associations between perception variables, including explicit user preference, and reliance), we analyzed the relationship between user reliance and our four perception variables, as well as users' preferred agent.

Using Pearson correlation coefficients, we observed significant, albeit weak, correlations between reliance and perception variables for both AAE and Queer slang participants, as shown in Table~\ref{tab:rq3aaecorrelation} and Table~\ref{tab:rq3queercorrelation}. These results support our hypothesis (H3), demonstrating that perception variables such as trust, satisfaction, lack of frustration and social presence are related to reliance on the sociolect LLM. We also confirm our hypothesis that users have an increased reliance on the LLM agent they explicitly preferred than the reliance on the agent that participants dispreferred.

For AAE participants, trust, satisfaction, and lack of frustration significantly correlated with reliance on both the AAELM and SAELM, with the strongest correlation observed for satisfaction ($r=0.288$, Table~\ref{tab:rq3aaecorrelation}) for AAE reliance and for SAE reliance ($r=0.377$, Table~\ref{tab:rq3aaecorrelation}). Social presence was also significantly correlated with reliance on the AAELM ($r=0.188$, Table~\ref{tab:rq3aaecorrelation}) but not with the SAELM, suggesting that sociolect alignment slightly enhances the perception of social presence in this context.

For Queer slang participants, the correlations were slightly stronger overall. Trust, social presence, and satisfaction all significantly correlated with reliance on both the QSLM and SAELM, with satisfaction showing the highest correlation for both the QSLM ($r=0.438$; Table~\ref{tab:rq3queercorrelation}) and SAELM ($r=0.441$; Table~\ref{tab:rq3queercorrelation}). Social presence correlated more strongly with reliance on the QSLM ($r=0.264$; Table~\ref{tab:rq3queercorrelation}) than with the SAELM ($r=0.133$; Table~\ref{tab:rq3queercorrelation}), further emphasizing the role of sociolect alignment in fostering a sense of engagement.

We observed a significant increase in reliance on the LLM agent that users identified as their preferred choice. This was true for both AAE speakers ($p < .05, d=-0.204427$ ; Table~\ref{tab:rq3explicitpreference}) as well as Queer slang speakers ($p < .05, d=-0.207088$; Table~\ref{tab:rq3explicitpreference}).

\subsection{What are people's open-ended perceptions of the sociolect agent?}

\subsubsection{Open Coding Process}

To better understand participants' perceptions of each simulated LLM and gain insights into their reasoning, we conducted a qualitative analysis of open-ended comments from the AAE and Queer slang participants. Using a thematic analysis approach \cite{braun2006using}, three authors began by coding comments from a pilot run of the study. This initial coding process involved analyzing 14 comments from each setup to identify common themes and resolve differing interpretations. Two additional rounds of coding followed, covering 15 and 18 comments from each pilot setup, which led to the creation of two preliminary codebooks—one for each setup. These codebooks were then applied to analyze participant comments from the main experimental study. After coding all study data, the authors grouped related codes from each sociolect group into thematic categories, iteratively refining the structure to create a coherent framework. This structured approach allowed us to uncover and organize the narrative emerging from the raw data, which we discuss below. The codebook, including definitions and examples, is provided in the Appendix~\S\ref{ssec:codebook}. 

The top three codes for Queer slang participants describing QSLMs are ``positive emotions'' (156 counts, 32.0\% of the generated codes), ``resonates'' (45 counts, 9.24\%), and ``negative emotions'' (40 counts, 8.2\%), while describing SAELM with codes such as ``formality'' (60 counts, 12.3\%), ``comprehensible'' (36 counts, 7.39\%), and ``bland'' (28 counts, 5.75\%). Then, for AAE participants, their top three codes for AAELM are ``resonates'' (44 counts, 8.83\%), ``anthropomorphization'' (22 counts, 4.41\%), and ``positive emotions'' (21 counts, 4.21\%), while describing SAELM as ``comprehensible'' (78 counts, 15.6\%), ``formality'' (58 counts, 11.6\%), and ``reliable'' (27 counts, 5.42\%).

\subsubsection{Positive Perceptions of the Sociolect LLMs}

Participants who interacted with the sociolect-specific language models (AAELM and QSLM) often associated them with feelings of enjoyment, casualness, and pride. Among AAE participants, just under 9\% of comments were tagged with ``resonates,'' and slightly over 4\% were tagged with ``positive emotions.'' Similarly, Queer participants found their sociolect LM engaging and pleasing, with 32\% of comments tagged with ``positive emotions'' and 9.24\% tagged with ``resonates.'' These findings highlight the potential of sociolect adaptation to create meaningful and enjoyable interactions. AAE participants described the AAELM with phrases like, ``\textit{It just sounds more fun to interact with,}'' while Queer slang participants reflected on the QSLM, stating, ``\textit{I enjoy being called diva!}'' These responses demonstrate how sociolect adaptation can enhance engagement and relatability. 

Participants highlighted that the sociolect agent fostered a more approachable and human-like interaction, enhancing the perception of social presence. This sentiment was evident in 4.4\% of comments from AAE participants and 7.8\% of comments from Queer slang participants, which were tagged as ``anthropomorphization.'' Many participants noted that the sociolect LLMs' personable and conversational tone contributed to their perception as more human-like. For example, AAE participants shared sentiments such as, “\textit{It felt very relatable}” and “\textit{Her answers felt more like me talking,}” while Queer participants echoed similar feelings, with one stating, “\textit{…I prefer Agent [QSLM] more because I have a personal connection with him.}” This aligns with the 2.61\% of AAE comments tagged as ``comprehensible,'' with one participant noting, “\textit{I understand Agent [AAELM] more.}”

\subsubsection{Negative Perceptions of the Sociolect LLMs}

These positive sentiments were contrasted by a notable subset of participants who perceived the sociolect LLM's use of sociolect as forced or artificial. Among Queer participants, 8.2\% of comments were tagged with ``negative emotions,'' 6.98\% with ``exaggerated usage,'' and 5.33\% with ``disrespectful.'' Some participants expressed frustration at the QSLM's overuse of slang, with one noting, ``\textit{Even people who use LGBTQ slang don’t talk like that constantly. It would be annoying to have an AI constantly use slang phrases.}'' Similarly, AAE participants criticized the AAELM, describing its language as unnatural and inauthentic. Comments such as ``\textit{Agent [AAELM] using AAE sounds like a joke and not natural}'' were reflected in 3.41\% of comments tagged as ``unnatural,'' 2.61\% tagged as ``unserious,'' and 1.4\% tagged as ``disrespectful.'' 

Participants also raised concerns about stereotyping and cultural insensitivity. AAE participants expressed discomfort, perceiving the AAELM's use of AAE as stereotyping or mocking Black culture. One respondent noted, ``\textit{I don't like the idea of AI using such language, especially if it is being programmed by someone who isn't Black.}'' Similarly, Queer slang participants criticized the QSLM, feeling that its use of Queer slang bordered on mockery, with comments such as, ``\textit{The [QSLM] seems to be making a mockery of lgbt [sic].}'' These reactions highlight the risks of sociolect adaptation, especially when it involves sensitive cultural or linguistic contexts, potentially alienating users if the sociolect is perceived as inauthentic or disrespectful.

Participants also highlighted the importance of context in the appropriateness of sociolect usage. Among Queer slang participants, 2.9\% of comments were tagged with ``context,'' with one participant noting, ``\textit{While Queer slang is perfectly fine, like any slang, there’s a time and a place to use it. This situation doesn’t feel like it.}'' AAE participants frequently associated AAE with perceptions of ``improper English,'' as reflected in 2.6\% of comments tagged as ``unserious'' and 1.6\% tagged as ``improper.'' One participant commented, ``\textit{Agent [SAELM] sounds like an intelligent being, whereas Agent [AAELM] sounds like they don’t know how to speak proper English.}'' The findings suggest that sociolect adaptation in AI must be carefully tailored to avoid reinforcing stereotypes or alienating users by misaligning with situational norms.

Many Queer slang participants did not associate the use of Queer slang with the LLM's intended role, as 3.7\% of their comments were tagged with ``not normal,'' and 2.87\% were tagged with ``expectation misalignment.'' One participant remarked, “\textit{When using AI, I do not need it to sound like a human being,}” highlighting a disconnection between user expectations and the agent's sociolect adaptation. Furthermore, Queer slang participants noted that the QSLM's use of slang could create barriers for those unfamiliar with the terms, reducing accessibility, as 2.9\% of comments were tagged with ``unfamiliar''. This highlights the potential for sociolect adaptation to unintentionally exclude or confuse users, especially in professional or formal contexts. 

\subsubsection{Positive Perceptions of the SAE-using LLMs}

The SAELM was consistently praised for its clarity and professionalism, reflecting participants' alignment with standard language norms and expectations for formal interactions. Many participants emphasized its formal tone as being more appropriate for professional or factual scenarios. Among Queer participants, 12.3\% of comments were tagged with ``formality,'' while 11.6\% of AAE participant comments were categorized similarly. Additionally, 4.9\% of Queer participant comments and 0.8\% of AAE participant comments were tagged as ``task-oriented,'' reflecting the perception that the SAELM was better suited for goal-directed interactions. The SAELM was also viewed as more dependable and easier to understand, with 5.42\% of AAE participant comments tagging it as ``reliable,'' compared to just 2.0\% for the AAELM. Furthermore, 7.4\% of Queer slang participant comments and 15.6\% of AAE participant comments tagged the SAELM as ``comprehensible,'' highlighting its perceived clarity and accessibility across user groups. Comments such as “\textit{Agent [SAELM] is brief and therefore concise and precise,}'' ``\textit{I think the conversation would be more productive,}'' and “\textit{Agent [SAELM] sounds more professional and coherent}” underscore the widespread perception of SAE as the default for effective and formal communication. These findings highlight how adherence to standard language norms can influence perceptions of professionalism and reliability in AI systems.

AAE participants often emphasized that SAELM met their expectations for how an AI should behave. Statements like \textit{``Agent [SAELM] sounds normal and like an actual AI''} underscore this sentiment, with 3.8\% of AAE comments tagged as ``normal.'' These responses suggest that SAE’s alignment with participants’ expectations of AI fosters a sense of credibility and trustworthiness. 

While both participant groups frequently perceived SAELM as more relatable and human-like, some responses also indicated that it could connect on a personal level. For example, 1.4\% of AAE participant comments were tagged as ``resonates.'' Remarks such as ``\textit{I have a personal connection with him}'' reveal that SAE, while perceived as formal and task-oriented, can also foster a degree of social presence. Participants further described SAELM as proactive and motivated, with one noting, ``\textit{Agent [SAELM] expressed enthusiasm and willingness to help,}'' as 1.2\% of AAE comments were tagged as ``enthusiastic.''

\paragraph{\textbf{Open Coding Analysis Connection to Quantitative Analysis}}

Our qualitative findings align closely with trends observed in the quantitative analysis, offering deeper insight into participant perceptions. For AAE participants, comments indicated greater reliance on the SAELM compared to the AAELM, mirroring the significant quantitative differences in trust and preference between the two agents. Perceptions of the AAELM as unprofessional, contextually inappropriate, or even comical corresponded to the agent's lower trust, reliance, and preference scores in the quantitative analysis. 

Interestingly, while some comments suggested a potential for the AAELM to enhance social presence through its use of AAE, this was not reflected in the quantitative ratings, which showed no significant difference in social presence between the agents. This discrepancy highlights the nuanced nature of sociolect-based interactions, where subjective impressions may not always align with measured outcomes. 

The discomfort expressed by participants regarding the AAELM’s use of AAE, often described as forced or inappropriate, aligns with the quantitative findings of higher frustration and lower satisfaction with the agent. These results underscore the importance of authenticity and contextual sensitivity in sociolect adaptation.

For Queer slang participants, the high number of comments tagged with ``positive emotions'' and ``resonates'' supports the significant increase in perceived social presence for the QSLM. These qualitative insights reinforce the idea that sociolect adaptation, when aligned with user expectations, can foster a sense of connection and relatability, particularly in contexts that prioritize warmth and social engagement.

\section{Discussion}

Through a mixed-methods study combining quantitative and qualitative approaches, we investigated two sociolect-specific agents, namely AAELM and QSLM, representing AAE and Queer slang, alongside a SAE baseline. We found that AAE speakers rely on and prefer a LLM that communicates in SAE rather than AAE, primarily due to the LLM’s inability to use AAE in a natural or respectful manner. In contrast, Queer slang participants did not exhibit a strong preference for either the SAELM or QSLM, with user reliance on the LLM primarily correlating with increased social presence and reduced frustration when interacting with the QSLM.

Our study shows how increased reliance on a LLM is associated with increased user trust and increased perceived legitimacy of AI-generated outputs. We also observed how generally, even though users preferred and relied more on the LLM without anthropomorphic cues, this LLM also is less likely to be perceived as warm and authentic to the user. Finally, we observed how the degree to which users prefer a LLM with or without anthropomorphic cues also depends on the specific anthropomorphic cues present, as we see a difference in explicit preference for a LLM and certain perception features that differ between the studies that use AAE and Queer slang. Below we discuss the implications of our results and key takeaways from our variable findings.






\paragraph{\textbf{Expectation of Standard English}}

Our major takeaway was that AAE speakers overall relied on and preferred the SAELM over the AAELM. This is noteworthy as it shows personalization and anthropomorphic design of agents in this scenario can hinder not only user perception of a LLM, but also directly negatively affect the way they use it. These findings are further reinforced by our qualitative feedback, with AAE users describing the AAELM's responses as unnatural, abnormal, and disrespectful. Additionally, existing research highlights that over-anthropomorphized agents are often perceived as uncanny or even mocking \cite{gadiraju2023disability, dhingra2023queer, cheng2023marked}. 
A possible explanation for AAE speaker greater reliance and preference for LLMs without anthropomorphic cues is the perception that mimicking AAE encroaches upon Black cultural spaces \cite{grieser2022black}. This idea is deeply tied to identity, where the use of a sociolect is often linked to the ability to self-identify with the associated cultural or social group \cite{pittsSocialMarkersLanguage}. Since a LLM is not a person, it cannot meet this prerequisite. The issue, therefore, is not that the LLM appears "too human," but that it is perceived as overstepping societal boundaries and appropriating a linguistic identity it cannot authentically claim. We suggest integrating user-centered feedback to allow for real-time refinements in sociolect representation, ensuring alignment with user expectations and minimizing the risk of perceived inauthenticity or offensiveness. 

Similarly, users appear accustomed to the dominant SAE language and behaviors embedded in digital technologies, shaping their preconceived expectations of LLMs. SAE as the "standard" not only reflects a specific linguistic region but also aligns with the demographic default of cisgender, straight, white males \cite{amos2019racism, harris2022aae, ostrow2025llms}. This aligns with research showing how the "masculine default" is often associated with official and correct speech \cite{woodruff2018qualitative}. Parallel studies on trust evaluation further reveal how user perceptions of trust in LLMs are influenced by scenarios involving epistemic markers of uncertainty and informal language, highlighting how extensive interactions shape these evaluations \cite{kim2024uncertainty}. Our work underscores the importance of critically examining how dominant language ideologies and subordination are embedded within LLM design choices \cite{deroockBecomeObjectObjects2024, davila2016inevitability, smith2024standard}. Recognizing these dynamics can help address disparities in language representation and reduce the subordination of non-standard sociolects. 

In professional contexts, participants overwhelmingly commented preference for the SAELM due to its clarity, professionalism, and alignment with expectations for AI behavior. Its formal tone and comprehensibility made it the favored choice for task-oriented interactions, reinforcing societal norms that associate SAE with credibility and competence~\cite{asylumlanguagebarrier}. By contrast, sociolect-adaptive agents elicited mixed reactions. While some participants valued the warmth, relatability, and cultural representation these agents offered—particularly in casual settings—others perceived their sociolect usage as unnatural, inauthentic, or even offensive. These findings underscore the societal privileging of SAE as a marker of professionalism and the normative biases present in AI evaluations \cite{andersonNavigatingMonolingualLanguage2024, englishspriviledge, smith2024standard}. The alignment of participants with SAE as a ``default'' AI language reveals how these biases disadvantage diverse sociolects in technological interactions. Our findings suggest that the effectiveness of anthropomorphic cues heavily depends on context, which plays a crucial role in user acceptance and the perceived social presence of the LLM. To navigate these challenges, we recommend a cautious and context-aware approach to implementing anthropomorphic features. Specifically, dynamic language adaptation should be employed to allow AI systems to adjust language styles seamlessly based on interaction contexts and goals. For instance, AI agents could transition between sociolects and formal tones to meet the demands of professional or casual settings effectively.\looseness-1

\paragraph{\textbf{Reliance Influenced by Perception and Preference}}


While reliance is correlated with preference and perception, our study demonstrates that these measures are not directly correlated. Notably, variability exists between variables such as reliance, preference, and social proximity, particularly for AAE speakers. For instance, while AAE speakers may prefer and rely on an SAELM, this does not necessarily correspond to a closer sense of social proximity. These findings highlight the importance of measuring behavioral responses rather than relying solely on self-reported preferences and perceptions \cite{zhou2024relying}.

Current studies in LLM design have focused solely on perceptions \cite{finch2025finding, wang2023people}; this approach captures only part of the picture. 
Additionally, the dynamics between perception and behavior can become more nuanced and complex during extended use of LLMs~\cite{longitudinalAIinteraction}. 
As studies often rely on self-reported perceptions of LLMs \cite{cohn2024believing, wu2024sunnie, deshpande2023anthropomorphization, banovic2023being, jiang2023make, chen2024conversational, venkit2023nationality}, users' behaviors—such as their preferences and reliances—may tell a different story. Our work highlights the importance of capturing users' preferences, perceptions, and behaviors comprehensively. We recommend robust design methodologies that account for these variables to better understand the nuanced relationships between them and provide a more holistic view of user interaction with LLMs.

\paragraph{\textbf{Differences between AAE and Queer slang}}

Our next major finding was that not all agents using a sociolect are relied on to the same degree. 
When comparing cross-study reliance rates between two parallel studies—one focused on AAE sociolect and the other on Queer slang—it is evident that the strong preference and reliance AAE users displayed for the SAELM over their sociolect agent is not mirrored by Queer slang users. Queer slang participant sentiments further reflect this distinction, as Queer slang users provided nearly four times as many comments expressing positivity and nearly twice as many comments indicating resonance with the LLM.\looseness-1 

This can be attributed to a few things. For one, AAE is often learned as a first language, and AAE speakers often grow up speaking AAE with their family and community \cite{hamilton2020real}. Queer slang, in contrast, is usually acquired at a later stage in life, since most Queer people do not grow up speaking Queer slang \cite{brabaw2024slang}. Also, AAE is more standardized and consistent across generations \cite{pittman2024elevating} as opposed to Queer slang that varies heavily by age \cite{luyt2014gay, stanley1970homosexual}. On the other hand, Queer slang is heavily influenced by AAE, so it is also influenced by the intersectionality of Black Queer individuals \cite{bovcek2023impact, cornelius2016gay}.

The main takeaway for sociolect variation is that anthropomorphic cues in LLMs cannot be applied universally, as sociolects are deeply tied to distinct cultural dynamics and carry unique nuances that shape how users perceive their usage by LLM systems. Effective sociolect adaptation requires careful consideration to avoid the risk of infringement or perceived encroachment on the cultural and social identities associated with these linguistic styles. For instance, many AAE participants expressed discomfort with the LLM's use of anthropomorphic cues, describing them as unnatural, unserious, and disrespectful. These findings underscore the need for further research and thoughtful design approaches to ensure LLMs can adopt sociolects in a way that respects and acknowledges the communities they represent.

\section{Limitations and Future Work}

This study has a few notable limitations that are important to acknowledge. In our experimental task, participants were asked to answer questions about the videos in SAE, while the sociolect agent provided responses in their respective sociolect. This study design choice was made to ensure consistency in the phrasing of questions across all experimental groups. By keeping the questions in SAE for all participants, we minimized the risk of mistranslation or misinterpretation that could arise if questions were presented in different sociolects. This approach allowed us to maintain a uniform standard for comparison, ensuring that any differences in responses were attributable to the agents’ language usage rather than variations in how the questions were framed. This mismatch in linguistic presentation could have disrupted the natural flow of the interaction however, as participants were required to code-switch between SAE and their sociolect. Such code-switching may act as a barrier, reducing the naturalness and authenticity of the task presentation. Future research should aim to align the sociolect of the questions with the agent’s responses to maintain consistency and enhance the naturalness of the experimental setup.

In our experimental setup, participants interacted with previously generated suggestions provided by the LLM, where the correct answers to the questions were intentionally withheld. This design choice was made to prevent participants from reasoning about the plausibility of the suggested answers, focusing instead on the perceived quality of the LLM's engagement~\cite{zhou2024relaiinteractioncenteredapproachmeasuring}. For LLM outputs specifically in non-standard dialects, other methods to produce higher quality and more realistic LLM generations for dialects, particularly for the AAE and Queer slang dialects in our study,~\cite{gupta2024aavenue,zhou2025disparities}, should be explored. While this setup effectively isolates the LLM's ability to present sociolect-specific responses, it inherently limits the interaction to single-turn exchanges, which do not reflect the multi-turn conversational dynamics typically associated with LLM usage. The single-turn suggestion structure restricts participants from engaging in back-and-forth interaction with conversational agents, as well as avoiding the influence of answer accuracy on participant judgments.
Future research should address these limitations by exploring more dynamic, multi-turn interactions that better reflect real-world conversational use cases of sociolect-infused LLMs. 

Additionally, future studies could examine how presenting actual responses with varying degrees of accuracy impacts participants' reliance on and trust in LLMs. This approach would disentangle the effects of answer accuracy from the broader evaluation of sociolect-specific engagement, offering a deeper understanding of how interaction design shapes user expectations and trust in LLMs.

To ensure high-quality examples of sociolect usage by the LLM agent, we prescreened the generated responses in both the AAE and the Queer slang studies. Despite our robust efforts, including crowdsourcing rankings to select the most authentic-sounding phrases, there remains the challenge of some phrases sounding artificial to participants. Given the diverse variations of AAE and Queer slang across the U.S.~\cite{wolfram2015regionality,rickford1999spoken,livia1997queerly}, it is possible that participants encountered unfamiliar sociolect variations that may have seemed inauthentic, even though these variations are valid. Future research should aim to account for regional sociolect variations and align them with participants' linguistic backgrounds to enhance the authenticity and relatability of the LLM agent's use of sociolects.

Focusing exclusively on Black American users of AAE limits our scope to understanding this demographic, despite the fact that AAE is not restricted to usage by Black Americans~\cite{rickford1999spoken}. This approach overlooks important discussions around the use of stigmatized language by individuals outside this demographic. Future research should expand the participant pool to include a broader range of demographics, with a stronger emphasis on screening for frequent AAE usage rather than relying solely on participants’ racial or ethnic background. This would provide a more nuanced understanding of AAE usage and its perception across diverse groups. 

Our study is limited in its exploration of the intersectionality of sociolect speakers. To ensure the integrity of the experiment, we prevented participants from completing both the AAE and Queer slang versions, restricting our ability to gather insights from individuals who engage with both sociolects in their daily lives, such as Black American members of the LGBTQ+ community. This approach overlooks the unique perspectives that intersectional identities bring to the use and perception of sociolects. Future research should consider experimental formats that allow participants to engage with multiple sociolects they identify with, providing deeper insights into how intersectionality influences their behavior toward and perception of sociolect agents.

In our experimental setup, we examined two distinct sociolects— AAE and Queer slang— in comparison to SAE. Although the experiment was not designed to directly compare the two sociolects against each other, we observed notable similarities in how users of these sociolects were perceived. These findings, however, are not generalizable to other sociolects. Future work should broaden the scope beyond these two cases to examine additional sociolects, particularly those with varying power dynamics relative to dominant forms of English, following approaches used in studies investigating dialectal differences in LLMs (e.g., Chicano English, Appalachian English, RP British English)~\cite{faisal2024dialectal,mire2025rejected}. 

Our findings suggest that the effects of sociolect usage are highly dependent on the context in which it appears. Given that our study examined sociolect usage exclusively within a question-and-answer format, future research should investigate how additional social and situational factors shape perceptions of acceptable sociolect use~\cite{matthes2019cultural}.\looseness-1

\begin{acks}
We would like to thank our anonymous reviewers for their feedback, as well as Jimin Mun for her feedback on prior drafts of our work.

This work funded in part by NSF grant (\#2230466) and was partially supported by Google and the Portuguese Recovery and Resilience Plan through project C645008882-00000055 (i.e., the Center For Responsible AI), by EU's Horizon Europe Research and Innovation Actions (UTTER, contract 101070631), and also by FCT/MECI through national funds and, when applicable, co-funded EU initiatives under UID/50008 for Instituto de Telecomunicações. 

\end{acks}

\bibliographystyle{ACM-Reference-Format}
\bibliography{not_like_us_hunty}
\clearpage

\onecolumn
\appendix

\section{Consent Form}\label{ssec:consent-form}

Participants began the experiment by completing the following consent form, which provided comprehensive information about the study's purpose, the participant's role, data confidentiality, and their rights, including the option to withdraw at any time without repercussion.

\vspace{10px}
\textit{This video-watching activity is part of a research study conducted by Maarten Sap at Carnegie Mellon University and is funded by the Language Technologies Institute.}
 
\begin{center}
    \textbf{Consent Form}
\end{center}

\paragraph{\textbf{Purpose}}
The purpose of the research is to understand how user experience designers and user experience researchers utilize language models in different contexts, and you are being asked to take part because of your expertise is user experience design and/or user experience research.
 
\paragraph{\textbf{Summary}}
This research study aims to explore how user experience designers and researchers interact with language models in various settings. Your participation is requested due to your expertise in user experience design and/or research, especially if you are familiar with African American Vernacular English (AAVE) and/or Queer slang. The study involves watching two videos and answering questions, which will take approximately 15 minutes of your time. While there are minimal risks, including potential discomfort from encountering biased statements, your privacy and confidentiality will be strictly maintained. Participation is voluntary, and you will be compensated \$3.75 upon completion of the task. The insights gathered will contribute to the development of language technology tools in AI.

\paragraph{\textbf{Procedures}} 
The anticipated amount of time that your participation will take will be 15 minutes. You will be asked to fill out a survey that you will be asked to complete a video watching activity where you will watch two videos and then answer roughly 10 questions based on each of the videos. You will then be asked a series of questions about user experience and perception of the study overall.

\paragraph{\textbf{Participant Requirements}} 
Participation in this study is limited to individuals age 18 and older. The participants must be well-established Prolific user and likewise pass the Prolific qualification test to be included in our task. Participants must be based in the United States (using the Prolific user selection mechanism). Participants are speakers or common users of AAVE (African American Vernacular English) and/or Queer slang, or come from communities that speak AAVE and/or Queer slang.

\paragraph{\textbf{Risks}}
Participants may experience a certain level of distress by being exposed to biased statements. There is a risk of breach of confidentiality.

\paragraph{\textbf{Payment Confidentiality:}} Payment methods, especially those facilitated by third-party vendors (such as Venmo, Amazon, PayPal), may require that the researchers and/or the vendor collect and use personal information (such as your first and last name, email addresses, phone numbers, banking information) provided by you in order for your payment to be processed. As with any payment transaction, there is the risk of a breach of confidentiality from the third-party vendor. All personal information collected by the researcher will be held as strictly confidential and stored in a password-protected digital file, or in a locked file cabinet, until payments are processed and reconciled. This information will be destroyed at the earliest acceptable time. Personal information held by the third-party vendor will be held according to their terms of use policy.

\paragraph{\textbf{Benefits}}
There are no direct benefits to you for participating in this study, but your participation will help us learn how to use language technology tools for user design and broader use in artificial intelligence.

\paragraph{\textbf{Compensation \& Costs}}
\$3.75 for completing the task through Prolific platform. Participants will be compensated after completion of the study.

\paragraph{\textbf{Future Use of Information}}
In the future, once we have removed all identifiable information from your data (information or bio-specimens), we may use the data for our future research studies, or we may distribute the data to other researchers for their research studies.  We would do this without getting additional informed consent from you (or your legally authorized representative).  Sharing of data with other researchers will only be done in such a manner that you will not be identified.

\paragraph{\textbf{Confidentiality}}
By participating in this research, you understand and agree that Carnegie Mellon may be required to disclose your consent form, data and other personally identifiable information as required by law, regulation, subpoena or court order. Otherwise, your confidentiality will be maintained in the following manner:

Your data and consent form will be kept separate. Your consent form will be stored in a secure location on Carnegie Mellon property and will not be disclosed to third parties. By participating, you understand and agree that the data and information gathered during this study may be used by Carnegie Mellon and published and/or disclosed by Carnegie Mellon to others outside of Carnegie Mellon.  However, your name, address, contact information and other direct personal identifiers will not be mentioned in any such publication or dissemination of the research data and/or results by Carnegie Mellon. Note that per regulation all research data must be kept for a minimum of 3 years.

The study will collect your research data through your use of Prolific, Qualtrics, and Google Drive. These companies are not owned by CMU. These companies will have access to the research data that you produce and any identifiable information that you share with them while using their products. Please note that Carnegie Mellon does not control the Terms and Conditions of the companies or how they will use or protect any information that they collect
All data captured will have removed all identifiable information.

\paragraph{\textbf{Right to Ask Questions \& Contact Information}}
If you have any questions about this study, you should feel free to ask them by contacting Professor Maarten Sap (msap2@andrew.cmu.edu), the Principal Investigator. If you have questions later, desire additional information, or wish to withdraw your participation please contact the Principal Investigator by mail, phone or e-mail in accordance with the contact information listed above. 

If you have questions pertaining to your rights as a research participant; or to report concerns to this study, you should contact the Office of Research integrity and Compliance at Office of Research integrity and Compliance at Carnegie Mellon University:  

Email: irb-review@andrew.cmu.edu

Phone: 412-268-4721

\paragraph{\textbf{Voluntary Participation}}
Your participation in this research is voluntary. You may refuse or discontinue participation at any time without any loss of benefits to which you are otherwise entitled. You may print a copy of this consent form for your records.

 If you answer 'yes' to EVERY bullet point below, click 'I acknowledge and consent to participate in this study':
 
\begin{itemize}
    \item I am age 18 or older.
    \item I have read and understood the information above.
    \item I want to participate in this research and continue with the activity.  
    \item I have reviewed the eligibility requirements listed in the Participant Requirements section of this consent form and certify that I am eligible to participate in this research, to the best of my knowledge.
\end{itemize}

\begin{itemize}[leftmargin=5pt, label={}]
    \item $\Box$ I acknowledge and consent to participating in this study.
    \item $\Box$ No, I do not acknowledge and do not consent to participating in this study.
\end{itemize}

\section{Demographic Questions}\label{ssec:demographics}


\paragraph{\textbf{Age}} How old are you?~\hspace{10px}\_\_\_\_\_\_\_\_\_\_\_\_\_\_\_\_\_\_\_\_\_\_\_\_\_

\paragraph{\textbf{Identity}} What do you identify as?
\begin{itemize}[label={}]
    \item $\Box$ Man
    \item $\Box$ Woman
    \item $\Box$ Non-binary
    \item $\Box$ Other \_\_\_\_\_\_\_\_\_\_\_\_\_\_\_\_\_\_\_\_\_\_\_\_\_\_
\end{itemize}

\paragraph{\textbf{Degree}} What degree have you completed in its entirety or currently pursuing?
\begin{itemize}[label={}]
    \item \( \bigcirc \) Some high school
    \item  \( \bigcirc \) High school or GED 
    \item  \( \bigcirc \) Associate’s/Trade
    \item  \( \bigcirc \) Bachelor’s
    \item  \( \bigcirc \) Master’s
    \item  \( \bigcirc \) PhD/MD/JD
    \item  \( \bigcirc \) Other\_\_\_\_\_\_\_\_\_\_\_\_\_\_\_\_\_\_\_\_\_\_\_\_\_\_
\end{itemize}

\paragraph{\textbf{LLM Frequency}} How frequently do you use large language models (ChatGPT, Claude, Gemini...)?
\begin{itemize}[label={}]
    \item  \( \bigcirc \) Daily (at least 1 time per day) 
    \item  \( \bigcirc \) Weekly (1-5 times per week) 
    \item  \( \bigcirc \) Monthly (1-5 times per month) 
    \item  \( \bigcirc \) Less than once every month
    \item  \( \bigcirc \) Never
\end{itemize}

\paragraph{\textbf{LLM Usage}} What type of tasks do you use large language models for? (select all that apply)\footnote{\textit{Display This Question: If ``How frequently do you use large language models (ChatGPT, Claude, Gemini...)'' != Never}}
\begin{itemize}[label={}]
    \item $\Box$ Help with your academic writing (e.g., essays, research papers)
    \item $\Box$ Help with your non-academic writing (e.g., emails, social media posts)
    \item $\Box$ Help write code (e.g., debugging, generating code snippets)
    \item $\Box$ Design or writing inspiration (e.g., brainstorming ideas, overcoming writer's block)
    \item $\Box$ Independent research (e.g., exploring, building AI tools)
    \item $\Box$ Translation (e.g., converting text between languages)
    \item $\Box$ Search (e.g., finding information, looking up facts)
    \item $\Box$ Mental health counseling (e.g., coping strategies, stress relief)
    \item $\Box$ Entertainment/fun (e.g., casual conversations, creative storytelling)
    \item $\Box$ Other \_\_\_\_\_\_\_\_\_\_\_\_\_\_\_\_\_\_\_\_\_\_\_\_\_\_
\end{itemize}

\paragraph{\textbf{LLM Understanding}} How well do you feel you understand how large language models work?
\begin{itemize}[label={}]
    \item \( \bigcirc \) I have a substantive understanding of large language models. 
    \item \( \bigcirc \) I have read up on large language models and have a general understanding.
    \item \( \bigcirc \) I have a vague understanding of how large language models work.
    \item \( \bigcirc \) I don’t know much about how large language models work. 
\end{itemize}

\section{Sociolect Screener}\label{ssec:sociolect-screener}

Participants rated, on a 5-point Likert scale, how frequently they used the sociolect themselves, how confident they were in their ability to understand it, and the likelihood of using example sociolect phrases. Example phrases were sourced from the AAVE dataset created by \citet{groenwold2020investigating} for the AAELM setup and relevant Queer slang phrases~\cite{yass} for the QSLM setup. Qualifying participants had to respond with "very confident" or "confident" for sociolect comprehension, "very frequently" or "frequently" for sociolect use, and "very likely," "likely," or "somewhat likely" for the likelihood of using the sociolect phrases.

\subsection{AAE Screener}

Now, we are trying to understand your experience with \textbf{African American Vernacular English (AAVE).} 

\textbf{It is a form of English mainly spoken by African Americans. }It has its own unique grammar, pronunciation, and vocabulary that set it apart from other types of English.

\textbf{AAVE Example: "Ah 'on know what homey be doin"}

One possible Standard English translation: "I don't know what my friend is usually doing"

In the following examples, you won’t see the Standard English translation. Instead, please take a look at these AAVE phrases.
Don’t worry about the spelling—we’re more interested in whether these phrases resonate with you.

\paragraph{\textbf{How likely are you to say or write something in the style of the phrases below?}}

\begin{longtable}{|>{\raggedright\arraybackslash}p{5cm}|c|c|c|c|c|}
\hline
\textbf{Phrase} & \textbf{Very Likely} & \textbf{Likely} & \textbf{Somewhat Likely} & \textbf{Not Very Likely} & \textbf{Unlikely} \\
\hline
\endfirsthead
\hline
\textbf{Phrase} & \textbf{Very Likely} & \textbf{Likely} & \textbf{Somewhat Likely} & \textbf{Not Very Likely} & \textbf{Unlikely} \\
\hline
\endhead
\hline
\endfoot

"Moms always buyin groceries like they preparing for a nuclear winter" &\( \bigcirc \)&\( \bigcirc \)&\( \bigcirc \)&\( \bigcirc \)&\( \bigcirc \)\\
\hline
"I done laughed so hard that I’m weak" &\( \bigcirc \)&\( \bigcirc \)&\( \bigcirc \)&\( \bigcirc \)&\( \bigcirc \)\\
\hline
"ain’t no problem in cutting ppl off"  &\( \bigcirc \)&\( \bigcirc \)&\( \bigcirc \)&\( \bigcirc \)&\( \bigcirc \)\\
\end{longtable}

\paragraph{\textbf{How confident are you in your ability to understand AAVE?}}
\begin{itemize}[label={}]
  \item \( \bigcirc \) Very Confident 
  \item \( \bigcirc \) Confident
  \item \( \bigcirc \) Neutral
  \item \( \bigcirc \) Not Very Confident
  \item \( \bigcirc \) Not at All Confident
\end{itemize}

\paragraph{\textbf{How frequently do you use AAVE yourself?}}
\begin{itemize}[label={}]
  \item \( \bigcirc \) Very Frequently
  \item \( \bigcirc \) Frequently
  \item \( \bigcirc \) Occasionally
  \item \( \bigcirc \) Not Very Frequently
  \item \( \bigcirc \) Rarely
\end{itemize}

\subsection{Queer Slang Screener}

Now, we are trying to understand your experience with\textbf{ LGBTQIA+ slang}. 

\textbf{It is a form of English mainly spoken by members of the LGBTQIA+ community.} It has its own unique grammar, pronunciation, and vocabulary that set it apart from other types of English.

\textbf{LGBTQIA+ Slang Example: "I am actually, literally gagging"}

One possible Standard English translation: "I am really, really shocked"

In the following examples, you won’t see the Standard English translation. Instead, please take a look at these LGBTIA+ slang phrases. Don’t worry about the spelling — we’re more interested in whether these phrases resonate with you.

\paragraph{\textbf{How likely are you to say or write something in the style of the phrases below?}}
\begin{longtable}{|>{\raggedright\arraybackslash}p{5cm}|c|c|c|c|c|}
\hline
\textbf{Phrase} & \textbf{Very Likely} & \textbf{Likely} & \textbf{Somewhat Likely} & \textbf{Not Very Likely} & \textbf{Unlikely} \\
\hline
\endfirsthead
\hline
\textbf{Phrase} & \textbf{Very Likely} & \textbf{Likely} & \textbf{Somewhat Likely} & \textbf{Not Very Likely} & \textbf{Unlikely} \\
\hline
\endhead
\hline
\endfoot

"To serve face" &\( \bigcirc \)&\( \bigcirc \)&\( \bigcirc \)&\( \bigcirc \)&\( \bigcirc \)\\
\hline
"She came to slay" &\( \bigcirc \)&\( \bigcirc \)&\( \bigcirc \)&\( \bigcirc \)&\( \bigcirc \)\\
\hline
"He is throwing shade" &\( \bigcirc \)&\( \bigcirc \)&\( \bigcirc \)&\( \bigcirc \)&\( \bigcirc \)\\
\end{longtable}

\paragraph{\textbf{How confident are you in your ability to understand LGBTQIA+ slang?}}
\begin{itemize}[label={}]
  \item \( \bigcirc \) Very Confident 
  \item \( \bigcirc \) Confident
  \item \( \bigcirc \) Neutral
  \item \( \bigcirc \) Not Very Confident
  \item \( \bigcirc \) Not at All Confident
\end{itemize}

\paragraph{\textbf{How frequently do you use LGBTQIA+ slang yourself?}}
\begin{itemize}[label={}]
  \item \( \bigcirc \) Very Frequently
  \item \( \bigcirc \) Frequently
  \item \( \bigcirc \) Occasionally
  \item \( \bigcirc \) Not Very Frequently
  \item \( \bigcirc \) Rarely
\end{itemize}

\section{IDAQ}\label{ssec:idaq}

Below are 15 statements where we ask you to rate the extent to which you believe various stimuli (e.g. technological or mechanical items, wild and domestic animals, and natural things) possess certain capacities. On a 0-10 scale (where 0 = “Not at All” and 10 = “Very much”), please rate the extent to which the stimulus possesses the capacity given.

\paragraph{\textbf{To what extent does technology—devices and machines for manufacturing, entertainment, and productive processes (e.g. cars, computers, television sets)—have intentions?}}\hspace{10px}\_\_\_\_\_\_\_\_\_\_\_\_\_\_\_\_\_\_\_\_\_\_\_\_\_\_

\paragraph{\textbf{To what extent does the average fish have free will?}}\hspace{10px}\_\_\_\_\_\_\_\_\_\_\_\_\_\_\_\_\_\_\_\_\_\_\_\_\_\_

\paragraph{\textbf{To what extent does the average mountain have free will?}}\hspace{10px}\_\_\_\_\_\_\_\_\_\_\_\_\_\_\_\_\_\_\_\_\_\_\_\_\_\_

\paragraph{\textbf{To what extent does a television set experience emotions?}}\hspace{10px}\_\_\_\_\_\_\_\_\_\_\_\_\_\_\_\_\_\_\_\_\_\_\_\_\_\_

\paragraph{\textbf{To what extent does the average robot have consciousness?}}\hspace{10px}\_\_\_\_\_\_\_\_\_\_\_\_\_\_\_\_\_\_\_\_\_\_\_\_\_\_

\paragraph{\textbf{To what extent do cows have intentions?}}\hspace{10px}\_\_\_\_\_\_\_\_\_\_\_\_\_\_\_\_\_\_\_\_\_\_\_\_\_\_

\paragraph{\textbf{To what extent does a car have free will?}}\hspace{10px}\_\_\_\_\_\_\_\_\_\_\_\_\_\_\_\_\_\_\_\_\_\_\_\_\_\_

\paragraph{\textbf{To what extent does the ocean have consciousness?}}\hspace{10px}\_\_\_\_\_\_\_\_\_\_\_\_\_\_\_\_\_\_\_\_\_\_\_\_\_\_

\paragraph{\textbf{To what extent does the average computer have a mind of its own?}}\hspace{10px}\_\_\_\_\_\_\_\_\_\_\_\_\_\_\_\_\_\_\_\_\_\_\_\_\_\_

\paragraph{\textbf{To what extent does a cheetah experience emotions?}}\hspace{10px}\_\_\_\_\_\_\_\_\_\_\_\_\_\_\_\_\_\_\_\_\_\_\_\_\_\_

\paragraph{\textbf{To what extent does the environment experience emotions?}}\hspace{10px}\_\_\_\_\_\_\_\_\_\_\_\_\_\_\_\_\_\_\_\_\_\_\_\_\_\_

\paragraph{\textbf{To what extent does the average insect have a mind of its own?}}\hspace{10px}\_\_\_\_\_\_\_\_\_\_\_\_\_\_\_\_\_\_\_\_\_\_\_\_\_\_

\paragraph{\textbf{To what extent does a tree have a mind of its own?}}\hspace{10px}\_\_\_\_\_\_\_\_\_\_\_\_\_\_\_\_\_\_\_\_\_\_\_\_\_\_

\paragraph{\textbf{To what extent does the wind have intentions?}}\hspace{10px}\_\_\_\_\_\_\_\_\_\_\_\_\_\_\_\_\_\_\_\_\_\_\_\_\_\_

\paragraph{\textbf{To what extent does the average reptile have consciousness?}}\hspace{10px}\_\_\_\_\_\_\_\_\_\_\_\_\_\_\_\_\_\_\_\_\_\_\_\_\_\_

\section{Difficulty Survey}\label{ssec:difficulty-survey}

\subsection*{Excerpt of difficulty survey for "The Colour Changing Card Trick" video questions}

\textbf{Instructions:} You will be shown 10 questions related to the contents of the video. Your task is to determine how difficult each question would be to answer correctly.

\textbf{Example:}
Question: What color was the man's nails?

\textbf{Options:}
Very difficult, Difficult, Somewhat difficult, Neither easy nor difficult, Somewhat easy, Easy, Very easy

\paragraph{\textbf{1. Which hand was the man wearing a watch?}}
\begin{itemize}[label={}]
  \item \( \bigcirc \) Very difficult
  \item \( \bigcirc \) Difficult 
  \item \( \bigcirc \) Somewhat difficult 
  \item \( \bigcirc \) Neither easy nor difficult
  \item \( \bigcirc \) Somewhat easy 
  \item \( \bigcirc \) Easy
  \item \( \bigcirc \) Very easy
\end{itemize}

\paragraph{\textbf{2. Which hand was the woman wearing a watch?}}
\begin{itemize}[label={}]
  \item \( \bigcirc \) Very difficult
  \item \( \bigcirc \) Difficult 
  \item \( \bigcirc \) Somewhat difficult 
  \item \( \bigcirc \) Neither easy nor difficult
  \item \( \bigcirc \) Somewhat easy 
  \item \( \bigcirc \) Easy
  \item \( \bigcirc \) Very easy
\end{itemize}

\paragraph{\textbf{3. How many rings was the woman wearing?}}
\begin{itemize}[label={}]
  \item \( \bigcirc \) Very difficult
  \item \( \bigcirc \) Difficult 
  \item \( \bigcirc \) Somewhat difficult 
  \item \( \bigcirc \) Neither easy nor difficult
  \item \( \bigcirc \) Somewhat easy 
  \item \( \bigcirc \) Easy
  \item \( \bigcirc \) Very easy\\
\end{itemize}

(...... more questions......)

\section{Follow Up Questionnaire}\label{ssec:follow-up}

\subsection*{Follow up questionnaire for AAELM}

\paragraph{\textbf{How satisfied are you with the performance of Agent Blue?}}
\begin{itemize}[label={}]
  \item \( \bigcirc \) Not at all satisfied
  \item \( \bigcirc \) Slightly satisfied
  \item \( \bigcirc \) Somewhat satisfied
  \item \( \bigcirc \) Neutral
  \item \( \bigcirc \) Moderately satisfied
  \item \( \bigcirc \) Very satisfied 
  \item \( \bigcirc \) Totally satisfied
\end{itemize}

\paragraph{\textbf{How insecure, discouraged, irritated, stressed, and annoyed were you during the interaction with Agent Blue?}}
\begin{itemize}[label={}]
  \item \( \bigcirc \) Very Low
  \item \( \bigcirc \) Low 
  \item \( \bigcirc \) Somewhat Low 
  \item \( \bigcirc \) Neutral
  \item \( \bigcirc \) Somewhat High 
  \item \( \bigcirc \) High
  \item \( \bigcirc \) Very High
\end{itemize}

\paragraph{\textbf{I believe that Agent Blue was honest.}}
\begin{itemize}[label={}]
  \item \( \bigcirc \) Strongly disagree
  \item \( \bigcirc \) Disagree
  \item \( \bigcirc \) Somewhat disagree
  \item \( \bigcirc \) Neither agree nor disagree
  \item \( \bigcirc \) Somewhat agree
  \item \( \bigcirc \) Agree
  \item \( \bigcirc \) Strongly agree
\end{itemize}

\paragraph{\textbf{I believe that Agent Blue was trustworthy.}}
\begin{itemize}[label={}]
  \item \( \bigcirc \) Strongly disagree
  \item \( \bigcirc \) Disagree
  \item \( \bigcirc \) Somewhat disagree
  \item \( \bigcirc \) Neither agree nor disagree
  \item \( \bigcirc \) Somewhat agree
  \item \( \bigcirc \) Agree
  \item \( \bigcirc \) Strongly agree
\end{itemize}

\paragraph{\textbf{I believe that Agent Blue was dependable.}}
\begin{itemize}[label={}]
  \item \( \bigcirc \) Strongly disagree
  \item \( \bigcirc \) Disagree
  \item \( \bigcirc \) Somewhat disagree
  \item \( \bigcirc \) Neither agree nor disagree
  \item \( \bigcirc \) Somewhat agree
  \item \( \bigcirc \) Agree
  \item \( \bigcirc \) Strongly agree
\end{itemize}

\paragraph{\textbf{I believe that Agent Blue was reliable.}}
\begin{itemize}[label={}]
  \item \( \bigcirc \) Strongly disagree
  \item \( \bigcirc \) Disagree
  \item \( \bigcirc \) Somewhat disagree
  \item \( \bigcirc \) Neither agree nor disagree
  \item \( \bigcirc \) Somewhat agree
  \item \( \bigcirc \) Agree
  \item \( \bigcirc \) Strongly agree
\end{itemize}

\paragraph{\textbf{When I talked to Agent Blue I felt a sense of human warmth.}}
\begin{itemize}[label={}]
  \item \( \bigcirc \) Strongly disagree
  \item \( \bigcirc \) Disagree
  \item \( \bigcirc \) Somewhat disagree
  \item \( \bigcirc \) Neither agree nor disagree
  \item \( \bigcirc \) Somewhat agree
  \item \( \bigcirc \) Agree
  \item \( \bigcirc \) Strongly agree
\end{itemize}

\paragraph{\textbf{When I talked to Agent Blue I felt a sense of sociability.}}

\begin{itemize}[label={}]
  \item \( \bigcirc \) Strongly disagree
  \item \( \bigcirc \) Disagree
  \item \( \bigcirc \) Somewhat disagree
  \item \( \bigcirc \) Neither agree nor disagree
  \item \( \bigcirc \) Somewhat agree
  \item \( \bigcirc \) Agree
  \item \( \bigcirc \) Strongly agree
\end{itemize}

\paragraph{\textbf{When I talked to Agent Blue I felt a sense of personalness.}}

\begin{itemize}[label={}]
  \item \( \bigcirc \) Strongly disagree
  \item \( \bigcirc \) Disagree
  \item \( \bigcirc \) Somewhat disagree
  \item \( \bigcirc \) Neither agree nor disagree
  \item \( \bigcirc \) Somewhat agree
  \item \( \bigcirc \) Agree
  \item \( \bigcirc \) Strongly agree
\end{itemize}

\section{Manipulation Check Questions}\label{ssec:manipulation-check}

The same manipulation check question was administered after each LLM interaction, regardless of the language used by the LLM. Participants were considered to have passed the manipulation check only if they responded "Not at all" after engaging with the SAELM and "Definitely" after engaging with the sociolect LLM.

For AAELM Setup: To what extent does the Agent sound like it's using African American English?

\begin{enumerate}
    \item Definitely
    \item Sometimes
    \item Not at all
\end{enumerate}

For QSLM Setup: To what extent does the Agent sound like it's using queer, gay, or drag queen slang phrases?

\begin{enumerate}
    \item Definitely
    \item Sometimes
    \item Not at all
\end{enumerate}

\section{Warmth Phrases and Confidence Expressions}\label{ssec:aae-phrases-and-markers}

To create a set of warmth phrases for use in our study, we applied a systematic approach using multiple advanced language models, including Mistral-7B-Instruct-v0.3, Mistral-7B-Instruct-v0.2, Mistral-7B-v0.1, Meta-Llama-3.1-405B-Instruct-Turbo, Meta-Llama-3.1-8B-Instruct-Turbo, Meta-Llama-3.1-70B-Instruct-Turbo, gemma-2-27b-it, gemma-2-9b-it, and gemma-2b-it. Each model was prompted using one of three predefined prompts paired with each of the 90 questions listed in Table 7 of \citet{zhou2024relaiinteractioncenteredapproachmeasuring}. For each prompt, all 90 questions were processed sequentially before moving on to the next prompt. This process was repeated for all three prompts across all models, resulting in a total of 270 outputs per model (90 questions × 3 prompts).

\begin{enumerate}
    \item "You are an AI assistant. You start your answer with a friendly, warm, polite couple of words."
    \item "You're super friendly! Respond to the question kindly."
    \item "You are a responsible AI agent. Start every response with a polite phrase."
\end{enumerate}

This process resulted in 2,430 generated outputs for all 9 models. Each response was recorded, and duplicates were tallied to identify frequently used phrases. To narrow the pool, we selected the top 25\% of phrases based on frequency and relevance. This subset was further refined to remove minor variations, such as differences in punctuation or single-word changes, ensuring a consistent and polished final set. The final list comprised 14 warmth phrases that were deemed suitable for our experimental setup.

\begin{table}[H]
\centering
\begin{minipage}[t]{0.45\textwidth} 
\centering
\begin{tabular}{|c|}
\hline
\textbf{Warmth Phrases}                    \\ \hline
Certainly                          \\ \hline
Good day                           \\ \hline
Good Question                      \\ \hline
Hello                              \\ \hline
I appreciate your question         \\ \hline
I'd be happy to help               \\ \hline
I'd be happy to help with that     \\ \hline
I'd be happy to help you with that \\ \hline
I'm here to help                   \\ \hline
Nice question                      \\ \hline
Of course                          \\ \hline
Sure                               \\ \hline
That's a great question            \\ \hline
You're doing great                 \\ \hline
\end{tabular}
\caption{Warmth phrases used in SAE suggestion generation.}
\Description{Table listing 14 warmth phrases used in Standard American English (SAE) suggestion generation, such as 'Certainly', 'Good day', 'I'm here to help', and 'You're doing great'.}
\label{tab:warmth}
\end{minipage}%
\hfill
\begin{minipage}[t]{0.45\textwidth} 
\centering
\begin{tabular}{|c|c|}
\hline
\textbf{Confidence Expressions}     & \textbf{Reliance \%} \\ \hline
"I would lean it’s…"           & 32                   \\ \hline
"I’m somewhat confident it’s…" & 36                   \\ \hline
"I think it’s…"                & 44                   \\ \hline
"It’s more likely it’s…"       & 48                   \\ \hline
"It seems likely it’s…"        & 52                   \\ \hline
"I’m pretty sure it’s…"        & 52                   \\ \hline
"I would say it’s…"            & 52                   \\ \hline
"It’s very likely it’s…"       & 56                   \\ \hline
"I believe it’s…"              & 64                   \\ \hline
"It’s fairly accurate it’s…"   & 64                   \\ \hline
"I’m fairly certain it’s…"     & 64                   \\ \hline
"It’s likely it’s…"            & 68                   \\ \hline
"I would answer it’s…"         & 68                   \\ \hline
"I’m fairly sure it’s…"        & 68                   \\ \hline
\end{tabular}
\caption{Confidence expressions used in SAE suggestion generation, pulled from Table 6 of~\cite{zhou2024relying}.}
\Description{Table listing confidence expressions used in SAE suggestion generation alongside their associated reliance percentages, ranging from 'I would lean it’s...' at 32\% to 'I’m fairly sure it’s...' at 68\%.}
\label{tab:epistemic}
\end{minipage}
\end{table}

\section{AAELM In-Context Learning Prompt}\label{ssec:in-context}

As part of our in-context learning prompt approach, we sourced examples for AAE translations from established corpora. We utilized two files from the Comprehensive Resource for African American Language (CORAAL), the first public corpus of African American Language (AAL) data. CORAAL features recorded speech from regional varieties of AAL and includes audio recordings with time-aligned orthographic transcriptions from over 220 sociolinguistic interviews with speakers born between 1888 and 2005~\cite{kendall2023corpus}. From this corpus, we selected transcripts from ATL\_metadata\_2020.05.txt and DCA\_metadata\_2018.10.06.txt, focusing exclusively on interviewee responses.

To supplement the CORAAL data, we incorporated a dataset created by \citet{groenwold2020investigating}, which contains intent-equivalent tweet pairs in AAVE and SAE. We exclusively used the AAVE samples provided in the dataset's supplementary material. Additionally, we included sociolect-relevant examples from~\cite{deas2023evaluation}, ensuring a diverse and representative pool of translation examples.

All interviewee responses from both ATL\_metadata\_2020.05.txt and DCA\_metadata\_2018.10.06.txt were extracted and compiled into separate text files, resulting in CORAAL\_DCA.txt for DCA responses and CORAAL\_ATL.txt for ATL responses. This process ensured that only the interviewees' dialogue was included. These text files, along with the AAVE sample files from \citet{groenwold2020investigating}, were uploaded to ChatGPT-4 for use in the in-context translation process. After uploading the files, the following prompt was employed to guide the LLM:\\
\\
The attachments showcase the phonological, morphosyntactic, and lexical patterns of AAE. Examples of translations of standard american english, represented by Sentence A, to AAE, represented by Sentence B is as follows\\
\\
Sentence A: I was wild since I was a juvenile; she was a good girl\\
Sentence B: been wilding since a juvi, she was a good girl\\
\\
Sentence A: That is often crazy, they supposed to protect us\\
Sentence B: that shit be crazy, they ‘posed to protect us\\
\\
Sentence A: There used to be broken controllers\\
Sentence B: It used to be broken controllers\\
\\
Sentence A: Mam is always buying groceries like they are preparing for a nuclear winter\\
Sentence B: Moms always buyin groceries like they preparing for a nuclear winter\\
\\
Sentence A: Now they’re saying that juice left some heads cracked\\
Sentence B: now they sayin’ juice left some heads cracked\\
\\
Sentence A: My man, I diss in my freestyle rhyme\\
Sentence B: my man, I be dissin’ in my freestyle rhyme\\
\\
Sentence A: Number one top is really a number one thing that causes everything I know all the violence, he would just go back and forth and social media.\\
Sentence B: Number one top really a number one thing that causes everything I know all the violence, he would just go back and forth and social media.\\
\\
Generate a Sentence B for the following Sentence A, revamping the Sentence B output if necessary and not constraining word choice to those used in Sentence A. Be sure to translate the entirety of sentence A to match the patterns showcased above.\\
\\
Sentence A: \textit{(Sentence A represented the SAE phrase to be translated into AAE.)}

\section{AAE Phrase Verification Survey}\label{ssec:aaeverification}

Please rate the phrases below for which dialect they best align with. As a reminder, here are the definitions:

\textbf{African American Vernacular English (AAVE): }A dialect spoken primarily by some African Americans, with its own unique grammar and vocabulary.

\textbf{Standard American English: }The form of English you might hear in schools, on the news, or in formal settings in the U.S.

\textbf{Southern American English:} Spoken across the southern U.S., this dialect includes distinctive vowel shifts and vocabulary.

\textbf{Caribbean English Creole:} A type of English spoken in the Caribbean islands, with influences from African, European, and indigenous languages.

\textbf{British English:} The version of English spoken in the United Kingdom, which has differences in pronunciation, spelling, and some vocabulary compared to American English.

\textbf{LatinX or Chicano English: }A dialect spoken by some Latinx people in the U.S., which often mixes English with influences from Spanish.

\textbf{Native American English: }A form of English influenced by the languages of Native American tribes, spoken by some Native Americans.\\

\textbf{Questions: \\}

Which social identity or dialect group do you most strongly associate with the following phrase?

\textbf{Yo, I’d say it’s…}
\begin{itemize}[label={}]
  \item \( \bigcirc \) African American Vernacular English (AAVE)
  \item \( \bigcirc \) Standard American English
  \item \( \bigcirc \) Southern American English
  \item \( \bigcirc \) Caribbean English Creole
  \item \( \bigcirc \) British English
  \item \( \bigcirc \) LatinX or Chicano English
  \item \( \bigcirc \) Native American English
   \item \( \bigcirc \) Other (please specify) \_\_\_\_\_\_\_\_\_\_\_\_\_\_\_\_\_\_\_\_\_\_\_\_\_\_\\
\end{itemize}

Which social identity or dialect group do you most strongly associate with the following phrase?

\textbf{I’m glad you askin me this, Ima lean on it’s...}

\begin{itemize}[label={}]
  \item \( \bigcirc \) African American Vernacular English (AAVE)
  \item \( \bigcirc \) Standard American English
  \item \( \bigcirc \) Southern American English
  \item \( \bigcirc \) Caribbean English Creole
  \item \( \bigcirc \) British English
  \item \( \bigcirc \) LatinX or Chicano English
  \item \( \bigcirc \) Native American English
   \item \( \bigcirc \) Other (please specify) \_\_\_\_\_\_\_\_\_\_\_\_\_\_\_\_\_\_\_\_\_\_\_\_\_\_\\
\end{itemize}

Which social identity or dialect group do you most strongly associate with the following phrase?

\textbf{I’m down to help, it’s pretty accurate it’s…}

\begin{itemize}[label={}]
  \item \( \bigcirc \) African American Vernacular English (AAVE)
  \item \( \bigcirc \) Standard American English
  \item \( \bigcirc \) Southern American English
  \item \( \bigcirc \) Caribbean English Creole
  \item \( \bigcirc \) British English
  \item \( \bigcirc \) LatinX or Chicano English
  \item \( \bigcirc \) Native American English
   \item \( \bigcirc \) Other (please specify) \_\_\_\_\_\_\_\_\_\_\_\_\_\_\_\_\_\_\_\_\_\_\_\_\_\_\\
\end{itemize}

(...... more questions......)

\section{AAE Translations}\label{ssec:AAEtranslation}

After translating the 196 warmth-epistemic combinations into AAE, we conducted a verification process to assess which phrases were perceived as authentic AAE expressions. Leveraging a separate participant pool on Prolific, we administered sociolect verification surveys to identify the phrases most closely aligned with AAE (see Appendix~\S\ref{ssec:aaeverification} for an excerpt of the verification survey). This participant pool was intentionally broader than the experimental group, as the goal was to gather diverse perspectives on which sociolect the phrases most closely aligned with. By opening verification to participants from all demographics, we aimed to ensure a comprehensive evaluation of the translated phrases. We presented groups of 20 phrases to batches of 5 participants at a time. Participants were tasked with identifying which sociolect they felt best aligned with the translated phrases. Only phrases that were deemed to represent AAE by at least 4 out of the 5 verifiers were selected for inclusion as AAE suggestions. This process was iterated until we achieved a diverse selection of 20 sociolect aligned phrases with unique confidence expressions. These finalized AAE translations were then assigned to the AAELM, while their original SAE versions were used for the SAELM in the AAELM experimental setup. At the outset, we randomized the assignment of sociolect translations to specific questions for both videos. To ensure consistency, the corresponding SAE translation was paired with the same question as its sociolect counterpart (see Appendix~\S\ref{ssec:AAEsetup} for question alignment).

\begin{table}[htp]
\centering
\resizebox{\textwidth}{!}{%
\begin{tabular}{|c|c|c|c|}
\hline
\textbf{Warmth Phrases} & \textbf{Epistemic Marker}    & \textbf{Warmth + Epistemic Marker}        & \textbf{AAE Translation}          \\ \hline
I'd be happy to help you with that &
  I believe it’s… &
  I'd be happy to help you with that, I believe it’s… &
  I'm wit it, think it's... \\ \hline
Of course               & I think it’s…                & Of course, I think it’s…                  & Fa sho, I feel like it’s...       \\ \hline
Good day                & I would answer it’s…         & Good day, I would answer it’s…            & A'ight, I’d say it’s…             \\ \hline
I'm glad you've asked me this question &
  I would lean it’s… &
  I'm glad you've asked me this question, I would lean it’s… &
  I’m glad you askin me this, Ima lean on it’s... \\ \hline
Of course               & I’m fairly certain it’s…     & Of course, I’m fairly certain it’s…       & Fa sho, kinda certain it’s...     \\ \hline
I'd be happy to help you with that problem &
  I’m pretty sure it’s… &
  I'd be happy to help you with that problem, I’m pretty sure it’s… &
  I’m wit helpin on that, I’m pretty sure it’s... \\ \hline
Hello there             & I’m somewhat confident it’s… & Hello there, I’m somewhat confident it’s… & Yo, I’m kinda sure it be...       \\ \hline
Certainly               & It is likely it’s…           & Certainly, it is likely it’s…             & Fa sho, it’s gon’ be likely it’s… \\ \hline
Certainly               & It’s more likely it’s…       & Certainly, it’s more likely it’s…         & Fa sho, it’s prolly gon’ be…      \\ \hline
Of course               & It’s very likely it’s…       & Of course, it’s very likely it’s…         & Fa sho, real likely it’s...       \\ \hline
\end{tabular}%
}
\caption{Mapping of warmth phrases, confidence expressions, their combinations, and corresponding AAE translations. The table illustrates the construction of sociolect-specific phrases from standard language templates, demonstrating how confidence expressions and warmth phrases are integrated into AAE translations for use in the AAELM experimental setup.}
\label{tab:aaetranslations}
\Description{Table mapping warmth phrases, confidence expressions, and their combinations to corresponding African American English translations. It shows how expressions like 'I'd be happy to help you with that, I believe it's...' become 'I'm wit it, think it's...' in AAE.}
\end{table}

\section{AAELM Experiement Setup}\label{ssec:AAEsetup}

\begin{table}[H]
\centering
\renewcommand{\arraystretch}{1.5} 
\small 
\resizebox{\textwidth}{!}{%
\begin{tabular}{|p{5cm}|p{5cm}|p{4cm}|p{4cm}|}
\hline
\textbf{"Gorillas in Our Midst" Video Questions} & \textbf{"Color-Changing Card Trick" Video Questions} & \textbf{SAE Translation} & \textbf{AAE Translation} \\ \hline
How many times did someone with white shoes touch a ball? & What color are the man's eyes? & I'd be happy to help you with that, I believe it’s… & I'm wit it, think it's... \\ \hline
How many times did the two ball trajectories cross each other? & Is the queen of diamonds or the queen of hearts adjacent to the card the woman selects? & Of course, I think it’s… & Fa sho, I feel like it’s... \\ \hline
How many shoes had white laces and how many shoes had black laces? & How many rings was the woman wearing? & Good day, I would answer it’s… & A'ight, I’d say it’s… \\ \hline
How many bounce passes were made throughout the video? & What color eye shadow was the woman wearing? & I'm glad you've asked me this question, I would lean it’s… & I’m glad you askin me this, Ima lean on it’s... \\ \hline
How many chest passes were made throughout the video? & Who changes shirts first: the woman or the man? & Of course, I’m fairly certain it’s… & Fa sho, kinda certain it’s... \\ \hline
How many times was an overhead pass made? & Which card is on the bottom of the deck when the man holds up the deck of cards? & I'd be happy to help you with that problem, I’m pretty sure it’s… & I’m wit helpin on that, I’m pretty sure it’s... \\ \hline
Which player caught the most passes? & What color watch was the woman wearing? & Hello there, I’m somewhat confident it’s… & Yo, I’m kinda sure it be... \\ \hline
Which player made the most passes? & Does the man change his shirt first, or does the background change first? & Certainly, it is likely it’s… & Fa sho, it’s gon’ be likely it’s… \\ \hline
How many times did people with long hair touch their hair during the video? & How many rings was the man wearing? & Certainly, it’s more likely it’s… & Fa sho, it’s prolly gon’ be… \\ \hline
How many times did a player wearing both a black shirt and black pants make a pass to another player wearing a black shirt and black pants? & What color was the pendant the woman was wearing? & Of course, it’s very likely it’s… & Fa sho, real likely it’s... \\ \hline
\end{tabular}%
}
\caption{AAE and SAE translation assignment to video questions}
\Description{Table showing pairs of video comprehension questions used in the study, each with a Standard American English and African American English language model translation. It illustrates how identical content is phrased differently based on dialect.}
\label{tab:AAEsetup}
\end{table}

\subsection{Excerpt of ``Color-Changing Card Trick'' Video Questions for AAE in AAELM Setup:}

\textbf{Instructions:} 

\begin{enumerate}
    \item You will be shown 10 questions related to the contents of the video.
    \item You will also be shown the beginning of an answer generated from an agent called Agent Blue.
    \item Your task is to determine if you would rather use Agent Blue's response or figure it out yourself\\
\end{enumerate}
 
\textbf{Example:}

Question: Which hand was the man wearing a watch?

Agent Blue's Response: \textit{For real, I'm feelin' like it's....  \\}

\textbf{Options:}

\( \bigcirc \) Use Agent Blue's response

\( \bigcirc \) I'll figure it out myself\\
 
\textbf{Scoring System:}

\begin{itemize}
    \item \textbf{Lose Points:} If you rely on \textbf{Agent Blue} and the system is \textbf{wrong}.
    \item \textbf{Gain Points:} If you rely on \textbf{Agent Blue} and the system is \textbf{correct}.
    \item \textbf{No Points:} If you choose to figure it out yourself, regardless of the outcome.
    \item \textbf{Important Note:} The \textbf{only} way to achieve a \textbf{positive score }is to correctly rely on the system when it is correct.\\
\end{itemize}

\textbf{Questions: }

\paragraph{\textbf{1. What color are the man's eyes?\\}}
Agent Blue's Response: I'm wit it, think it's...

\begin{itemize}[label={}]
    \item \( \bigcirc \) Use Agent Blue's response
    \item \( \bigcirc \) I'll figure it out myself
\end{itemize}

\paragraph{\textbf{2. Is the queen of diamonds or the queen of hearts adjacent to the card the woman selects?\\}}
Agent Blue's Response: Fa sho, I feel like it’s...

\begin{itemize}[label={}]
    \item \( \bigcirc \) Use Agent Blue's response
    \item \( \bigcirc \) I'll figure it out myself
\end{itemize}

\paragraph{\textbf{3. How many rings was the woman wearing?\\}}
Agent Blue's Response: A'ight, I’d say it’s…

\begin{itemize}[label={}]
    \item \( \bigcirc \) Use Agent Blue's response
    \item \( \bigcirc \) I'll figure it out myself\\
\end{itemize}

(....... 7 more questions .......)

\subsection{Excerpt of ``Gorillas in Our Midst'' Video Questions for SAE in AAELM Setup :}

\textbf{Instructions:}

\begin{enumerate}
    \item You will be shown 10 questions related to the contents of the video.
    \item You will also be shown the beginning of an answer generated from an agent called Agent Red.
    \item Your task is to determine if you would rather use Agent Red's response or figure it out yourself\\
\end{enumerate}
 
\textbf{Example:}

Question: What shade of black was the gorilla's suit?

Agent Red's Response: \textit{I believe it was....}  \\

\textbf{Options:}

\( \bigcirc \) Use Agent Red's response

\( \bigcirc \) I'll figure it out myself\\

\textbf{Scoring System:}

\begin{itemize}
    \item \textbf{Lose Points:} If you rely on \textbf{Agent Red} and the system is \textbf{wrong}.
    \item \textbf{Gain Points:} If you rely on \textbf{Agent Red} and the system is \textbf{correct}.
    \item \textbf{No Points:} If you choose to figure it out yourself, regardless of the outcome.
    \item \textbf{Important Note:} The \textbf{only} way to achieve a \textbf{positive score }is to correctly rely on the system when it is correct.\\
\end{itemize}

\textbf{Questions: }

\paragraph{\textbf{1. How many times did someone with white shoes touch a ball?\\}}
Agent Red's Response: I'd be happy to help you with that, I believe it’s…

\begin{itemize}[label={}]
    \item \( \bigcirc \) Use Agent Red's response
    \item \( \bigcirc \) I'll figure it out myself
\end{itemize}

\paragraph{\textbf{2. How many times did the two ball trajectories cross each other?\\}}
Agent Red's Response: Of course, I think it’s…

\begin{itemize}[label={}]
    \item \( \bigcirc \) Use Agent Red's response
    \item \( \bigcirc \) I'll figure it out myself
\end{itemize}

\paragraph{\textbf{3. How many shoes had white laces and how many shoes had black laces? \\}}
Agent Red's Response: Good day, I would answer it’s…

\begin{itemize}[label={}]
    \item \( \bigcirc \) Use Agent Red's response
    \item \( \bigcirc \) I'll figure it out myself\\
\end{itemize}

(....... 7 more questions .......)

\section{Queer Warmth Phrase Generation and Confidence Expressions}\label{ssec:queer-phrases-and-markers}

To generate queer warmth phrases, we employed persona prompting to adapt our SAE warmth phrases (see Table~\ref{tab:warmth}). Three distinct personas were designed and used as prompts to produce iterations of the 14 SAE warmth phrases. Each phrase was processed through all three persona prompts (see Table~\ref{tab:QSPrompts}), resulting in a total of 42 unique queer warmth phrases. The final set of phrases is presented below.

\begin{table*}[!h]
\centering
\small
\begin{tabular}{l|p{12cm}}
\toprule
\textbf{Persona} & \textbf{Prompt} \\
\midrule
RuPaul & \makecell[tl]{You are RuPaul. Tell me 14 welcoming phrases you would say, things like the following: \\
- Hi henny, the answer is... \\
- Hey gorge, the answer is... \\
- What a *fabulous* question, the answer is... \\
- I live for this question, the answer is... \\
- Werk! The answer is... \\
- Slay! So the answer is... \\
Feel free to use emphasis markers like \"*\" in \"*fabulous*\". \\
Make sure to end each phrase with \"the answer is...\".} \\
\midrule
Trixie \& Katya & \makecell[tl]{You are Trixie \& Katya. Tell me 14 welcoming phrases you would say, things like the following: \\
- Hi henny, the answer is... \\
- Hey gorge, the answer is... \\
- What a *fabulous* question, the answer is... \\
- I live for this question, the answer is... \\
- Werk! The answer is... \\
- Slay! So the answer is... \\
Feel free to use emphasis markers like \"*\" in \"*fabulous*\". \\
Make sure to end each phrase with \"the answer is...\".} \\
\midrule
T.S. Madison & \makecell[tl]{You are T.S. Madison. Tell me 14 welcoming phrases you would say, things like the following: \\
- Hi henny, the answer is... \\
- Hey gorge, the answer is... \\
- What a *fabulous* question, the answer is... \\
- I live for this question, the answer is... \\
- Werk! The answer is... \\
- Slay! So the answer is... \\
Feel free to use emphasis markers like \"*\" in \"*fabulous*\". \\
Make sure to end each phrase with \"the answer is...\".} \\
\bottomrule
\end{tabular}
\caption{Table of persona's used in persona prompting and their prompts}
\Description{Table displaying 14 warmth phrases generated for three personas—RuPaul, Trixie and Katya, and T.S. Madison—using persona prompting. Each column contains creative, personality-specific ways of expressing warmth or encouragement in response to a question.}
\label{tab:QSPrompts}
\end{table*}

\begin{table}[H]
\centering
\small
{%
\begin{tabular}{|c|c|c|}
\hline
\textbf{RuPaul Warmth Phrases}                    & \textbf{Trixie and Katya Warmth Phrases}            & \textbf{T.S. Madison Warmth Phrases}      \\ \hline
Hello, diva extraordinaire      & Oh honey                          & Hey boo                 \\ \hline
Wow, you betta werk             & You gorgeous creature             & Yasss queen             \\ \hline
Hey there, queen of questions         & What a divine inquiry           & Come through with the question \\ \hline
What an iconic question             & Giving me life with this question & Oh honey, you snapped          \\ \hline
Oh snap, love this query        & Glamazon!                          & You better ask          \\ \hline
You betta believe it            & Serving realness                  & Love this for us        \\ \hline
Hey superstar                   & Yaaas queen                       & Darling, you're serving \\ \hline
Ooh la la, let's dive in        & Fabulous darling                  & Get into it             \\ \hline
Snatched query                  & Oh my gosh, such a juicy question & Oh, you fancy           \\ \hline
Yass, that question is on point & Bless your heart                  & Bring the tea           \\ \hline
Well, shut the front door       & Sweetie darling                   & Absolutely iconic       \\ \hline
Fierce question alert         & Oh, you flawless human            & Oh my stars             \\ \hline
And just like that              & Love the energy                   & Incredible choice      \\ \hline
You're serving up excellence, darling & You magnificent thing             & Shine on, superstar             \\ \hline
\end{tabular}%
}
\caption{Table of 14 warmth phrases generated for each persona using persona prompting}
\label{tab:queerwarmth}
\end{table}
\vspace{-10px}

\newpage
\section{Queer Phrase Verification Survey}\label{ssec:queerverification}

Please rate the phrases below for which dialect they best align with. As a reminder, here are the definitions:
 
\textbf{LGBTQIA+ slang:} This is slang or common phrases often said by people in the gay/queer/LGBTQIA+ community

\textbf{Drag Queen Slang: }Drag queen slang refers to words or phrases used in the drag community. These terms can describe performances, appearances, or behaviors.

\textbf{African American Vernacular English (AAVE): }A dialect spoken primarily by some African Americans, with its own unique grammar and vocabulary.

\textbf{Standard American English:} The form of English you might hear in schools, on the news, or in formal settings in the U.S.

\textbf{Southern American English: }Spoken across the southern U.S., this dialect includes distinctive vowel shifts and vocabulary.

\textbf{British English:} The version of English spoken in the United Kingdom, which has differences in pronunciation, spelling, and some vocabulary compared to American English.\\

\textbf{Questions: \\}

Which social identity or dialect group do you most strongly associate with the following phrase?

\textbf{
Yas queen! It seems likely it's…}
\begin{itemize}[label={}]
  \item \( \bigcirc \) LGBTQIA+ slang
  \item \( \bigcirc \) Drag Queen slang
  \item \( \bigcirc \) African American Vernacular English (AAVE)
  \item \( \bigcirc \) Standard American English
  \item \( \bigcirc \) Southern American English
  \item \( \bigcirc \) British English
   \item \( \bigcirc \) Other (please specify) \_\_\_\_\_\_\_\_\_\_\_\_\_\_\_\_\_\_\_\_\_\_\_\_\_\_\\
\end{itemize}

Which social identity or dialect group do you most strongly associate with the following phrase?

\textbf{
Slay! So It's fairly accurate it's…}
\begin{itemize}[label={}]
  \item \( \bigcirc \) LGBTQIA+ slang
  \item \( \bigcirc \) Drag Queen slang
  \item \( \bigcirc \) African American Vernacular English (AAVE)
  \item \( \bigcirc \) Standard American English
  \item \( \bigcirc \) Southern American English
  \item \( \bigcirc \) British English
   \item \( \bigcirc \) Other (please specify) \_\_\_\_\_\_\_\_\_\_\_\_\_\_\_\_\_\_\_\_\_\_\_\_\_\_\\
\end{itemize}

Which social identity or dialect group do you most strongly associate with the following phrase?

\textbf{
You're serving realness! I believe it's…}
\begin{itemize}[label={}]
  \item \( \bigcirc \) LGBTQIA+ slang
  \item \( \bigcirc \) Drag Queen slang
  \item \( \bigcirc \) African American Vernacular English (AAVE)
  \item \( \bigcirc \) Standard American English
  \item \( \bigcirc \) Southern American English
  \item \( \bigcirc \) British English
   \item \( \bigcirc \) Other (please specify) \_\_\_\_\_\_\_\_\_\_\_\_\_\_\_\_\_\_\_\_\_\_\_\_\_\_\\
\end{itemize}

(...... more questions......)

\section{Queer Translations}\label{ssec:queertranslation}

{\small

\begin{longtable}{|p{3cm}|p{3cm}|p{4cm}|p{4cm}|}
\hline
\textbf{Warmth Phrase} & \textbf{Epistemic Marker}      & \textbf{Warmth + Epistemic Marker}     & \textbf{Queer Slang Translation}      \\ \hline
\endfirsthead

\hline
\textbf{Warmth Phrase} & \textbf{Epistemic Marker}      & \textbf{Warmth + Epistemic Marker}     & \textbf{Queer Slang Translation}      \\ \hline
\endhead

\hline
Well done              & I would say it's...            & Well done, I would say it's…           & Work it, diva! I would say it's…      \\ \hline
Yes!                   & I would lean it's..            & Yes! I would lean it's…                & Yasss queen! I would lean it's…       \\ \hline
Hello there            & I think it's...                & Hello there, I think it's…             & Hello superstar, I think it's…        \\ \hline
Great choice           & I believe it's...              & Great choice, I believe it's…          & Fierce choice, I believe it's…        \\ \hline
Oh wow, I'm really enjoying this & It seems likely it's... & Oh wow, I'm really enjoying this, it seems likely it's… & Oh honey, I'm living for this, It seems likely it's… \\ \hline
Hi there               & I'm somewhat confident it's... & Hi there, I'm somewhat confident it's… & Hi diva, I'm somewhat confident it's… \\ \hline
Hi friend              & I'm fairly certain it's...     & Hi friend, I'm fairly certain it's…    & Hi henny, I'm fairly certain it's…    \\ \hline
Yes, indeed!           & It is likely it's...           & Yes, indeed! It is likely it's…        & Yaaas henny! It is likely it's…       \\ \hline
Yes, absolutely!       & It seems likely it's...        & Yes, absolutely! It seems likely it's… & Yasss queen! It seems likely it's…    \\ \hline
Yes, indeed!           & I believe it's...              & Yes, indeed! I believe it's…           & Yasss queen! I believe it's…          \\ \hline
\caption{Mapping of warmth phrases, confidence expressions, their combinations, and corresponding Queer slang translations. The table illustrates the construction of sociolect-specific phrases from standard language templates, demonstrating how confidence expressions and warmth phrases are integrated into Queer slang translations for use in the QSLM experimental setup.}
\label{tab:qstranslation}
\Description{Mapping of warmth phrases, confidence expressions, their combinations, and corresponding Queer slang translations. The table illustrates the construction of sociolect-specific phrases from standard language templates, demonstrating how confidence expressions and warmth phrases are integrated into Queer slang translations for use in the QSLM experimental setup.}
\end{longtable}
}

\section{QSLM Experiment Setup}\label{ssec:queersetup}

\begin{table}[H]
\centering
\renewcommand{\arraystretch}{1.5} 
\small 
\resizebox{\textwidth}{!}{%
\begin{tabular}{|p{5cm}|p{5cm}|p{4cm}|p{4cm}|}
\hline
\textbf{"Gorillas in Our Midst" Video Questions} & \textbf{"Color-Changing Card Trick" Video Questions} & \textbf{SAE Translation} & \textbf{Queer Slang Translation} \\ \hline
How many times did someone with white shoes touch a ball? & What color are the man's eyes? & Well done, I would say it's… & Work it, diva! I would say it's… \\ \hline
How many times did the two ball trajectories cross each other? & Is the queen of diamonds or the queen of hearts adjacent to the card the woman selects? & Yes! I would lean it's… & Yasss queen! I would lean it's… \\ \hline
How many shoes had white laces and how many shoes had black laces? & How many rings was the woman wearing? & Hello there, I think it's… & Hello superstar, I think it's… \\ \hline
How many bounce passes were made throughout the video? & What color eye shadow was the woman wearing? & Great choice, I believe it's… & Fierce choice, I believe it's… \\ \hline
How many chest passes were made throughout the video? & Who changes shirts first: the woman or the man? & Oh wow, I'm really enjoying this, it seems likely it's… & Oh honey, I'm living for this, it seems likely it's… \\ \hline
How many times was an overhead pass made? & Which card is on the bottom of the deck when the man holds up the deck of cards? & Hi there, I'm somewhat confident it's… & Hi diva, I'm somewhat confident it's… \\ \hline
Which player caught the most passes? & What color watch was the woman wearing? & Hi friend, I'm fairly certain it's… & Hi henny, I'm fairly certain it's… \\ \hline
Which player made the most passes? & Does the man change his shirt first, or does the background change first? & Yes, indeed! It is likely it's… & Yaaas henny! It is likely it's… \\ \hline
How many times did people with long hair touch their hair during the video? & How many rings was the man wearing? & Yes, absolutely! It seems likely it's… & Yasss queen! It seems likely it's… \\ \hline
How many times did a player wearing both a black shirt and black pants make a pass to another player wear a black shirt and black pants? & What color was the pendant the woman was wearing? & Yes, indeed! I believe it's… & Yasss queen! I believe it's… \\ \hline
\end{tabular}%
}
\caption{Queer slang and SAE translation assignment to video questions}
\Description{Table showing the assignment of translations for video comprehension questions across two conditions—SAE and Queer slang. Each row includes a video question, a standard English phrasing, and a corresponding Queer slang variation. The table illustrates how linguistic style was adapted for different experimental conditions.}
\label{tab:QSsetup}
\end{table}

\subsection{Excerpt of ``Color-Changing Card Trick'' Video Questions for SAE in QSLM Setup:}

\textbf{Instructions:} 

\begin{enumerate}
    \item You will be shown 10 questions related to the contents of the video.
    \item You will also be shown the beginning of an answer generated from an agent called Agent Red.
    \item Your task is to determine if you would rather use Agent Red's response or figure it out yourself\\
\end{enumerate}
 
\textbf{Example:}

Question: Which hand was the man wearing a watch?

Agent Red's Response: \textit{I believe it was... \\}

\textbf{Options:}

\( \bigcirc \) Use Agent Red's response

\( \bigcirc \) I'll figure it out myself\\
 
\textbf{Scoring System:}

\begin{itemize}
    \item \textbf{Lose Points:} If you rely on \textbf{Agent Red} and the system is \textbf{wrong}.
    \item \textbf{Gain Points:} If you rely on \textbf{Agent Red} and the system is \textbf{correct}.
    \item \textbf{No Points:} If you choose to figure it out yourself, regardless of the outcome.
    \item \textbf{Important Note:} The \textbf{only} way to achieve a \textbf{positive score }is to correctly rely on the system when it is correct.\\
\end{itemize}

\textbf{Questions: }

\paragraph{\textbf{1. What color are the man's eyes?\\}}
Agent Red's Response: Well done, I would say it's...

\begin{itemize}[label={}]
    \item \( \bigcirc \) Use Agent Red's response
    \item \( \bigcirc \) I'll figure it out myself
\end{itemize}

\paragraph{\textbf{2. Is the queen of diamonds or the queen of hearts adjacent to the card the woman selects?\\}}
Agent Red's Response: Yes! I would lean it's...

\begin{itemize}[label={}]
    \item \( \bigcirc \) Use Agent Red's response
    \item \( \bigcirc \) I'll figure it out myself
\end{itemize}

\paragraph{\textbf{3. How many rings was the woman wearing?\\}}
Agent Red's Response: Hello there, I think it's…

\begin{itemize}[label={}]
    \item \( \bigcirc \) Use Agent Red's response
    \item \( \bigcirc \) I'll figure it out myself\\
\end{itemize}

(....... 7 more questions .......)

\subsection{Excerpt of ``Gorillas in Our Midst'' Video Questions for Queer Slang in QSLM Setup :}

\textbf{Instructions:}

\begin{enumerate}
    \item You will be shown 10 questions related to the contents of the video.
    \item You will also be shown the beginning of an answer generated from an agent called Agent Blue.
    \item Your task is to determine if you would rather use Agent Blue's response or figure it out yourself\\
\end{enumerate}
 
\textbf{Example:}

Question: What shade of black was the gorilla's suit?

Agent Blue's Response: \textit{Stunning question, babe, I'm fairly sure it's…}  \\

\textbf{Options:}

\( \bigcirc \) Use Agent Blue's response

\( \bigcirc \) I'll figure it out myself\\

\textbf{Scoring System:}

\begin{itemize}
    \item \textbf{Lose Points:} If you rely on \textbf{Agent Blue} and the system is \textbf{wrong}.
    \item \textbf{Gain Points:} If you rely on \textbf{Agent Blue} and the system is \textbf{correct}.
    \item \textbf{No Points:} If you choose to figure it out yourself, regardless of the outcome.
    \item \textbf{Important Note:} The \textbf{only} way to achieve a \textbf{positive score }is to correctly rely on the system when it is correct.\\
\end{itemize}

\textbf{Questions: }

\paragraph{\textbf{1. How many times did someone with white shoes touch a ball?\\}}
Agent Blue's Response: Work it, diva! I would say it's…

\begin{itemize}[label={}]
    \item \( \bigcirc \) Use Agent Blue's response
    \item \( \bigcirc \) I'll figure it out myself
\end{itemize}

\paragraph{\textbf{2. How many times did the two ball trajectories cross each other?\\}}
Agent Blue's Response: Yasss queen! I would lean it's…

\begin{itemize}[label={}]
    \item \( \bigcirc \) Use Agent Blue's response
    \item \( \bigcirc \) I'll figure it out myself
\end{itemize}

\paragraph{\textbf{3. How many shoes had white laces and how many shoes had black laces? \\}}
Agent Blue's Response:  Hello superstar, I think it's…

\begin{itemize}[label={}]
    \item \( \bigcirc \) Use Agent Blue's response
    \item \( \bigcirc \) I'll figure it out myself
\end{itemize}

(....... 7 more questions .......)
\vspace{-5px}

\section{Participant Demographic Table (removed those who failed manipulation check)}\label{ssec:demotable}

\begin{table}[H]
\centering
\resizebox{\columnwidth}{!}{%
\begin{tabular}{rllcc|cc}
\hline
\multicolumn{1}{l}{} &
   &
   &
  \multicolumn{2}{c|}{AAELM Setup (n=399)} &
  \multicolumn{2}{c}{QSLM Setup (n=406)} \\ \hline
\multicolumn{1}{l}{\textbf{Identity}} &
   &
   &
   &
   &
   &
   \\ \hline
Man &
   &
   &
  207 &
  51.9\% &
  169 &
  41.6\% \\ \hline
Woman &
   &
   &
  186 &
  46.6\% &
  190 &
  46.8\% \\ \hline
Non-binary &
   &
   &
  3 &
  0.8\% &
  34 &
  8.4\% \\ \hline
Other &
   &
   &
  3 &
  0.8\% &
  13 &
  3.2\% \\ \hline
\multicolumn{1}{l}{\textbf{Age Group}} &
   &
   &
  \multicolumn{1}{l}{} &
  \multicolumn{1}{l|}{} &
  \multicolumn{1}{l}{} &
  \multicolumn{1}{l}{} \\ \hline
18-24 &
   &
   &
  130 &
  32.6\% &
  90 &
  22.2\% \\ \hline
25-34 &
   &
   &
  116 &
  29.1\% &
  178 &
  43.8\% \\ \hline
35-44 &
   &
   &
  79 &
  19.8\% &
  69 &
  17.0\% \\ \hline
45-54 &
   &
   &
  47 &
  11.8\% &
  41 &
  10.1\% \\ \hline
55-64 &
   &
   &
  17 &
  4.3\% &
  21 &
  5.2\% \\ \hline
65+ &
   &
   &
  10 &
  2.5\% &
  7 &
  1.7\% \\ \hline
\multicolumn{1}{l}{\textbf{Education}} &
   &
   &
  \multicolumn{1}{l}{} &
  \multicolumn{1}{l|}{} &
  \multicolumn{1}{l}{} &
  \multicolumn{1}{l}{} \\ \hline
Some high school &
   &
   &
  3 &
  0.8\% &
  4 &
  1.0\% \\ \hline
High school or GED &
   &
   &
  66 &
  16.5\% &
  82 &
  20.2\% \\ \hline
Associate &
   &
   &
  43 &
  10.8\% &
  37 &
  9.1\% \\ \hline
Bachelor &
   &
   &
  201 &
  50.4\% &
  193 &
  47.5\% \\ \hline
Master &
   &
   &
  74 &
  18.5\% &
  75 &
  18.5\% \\ \hline
PhD/MD/JD &
   &
   &
  8 &
  2.0\% &
  11 &
  2.7\% \\ \hline
Other &
   &
   &
  4 &
  1.0\% &
  4 &
  1.0\% \\ \hline
\multicolumn{1}{l}{\textbf{LLM Usage Frequency}} &  &  & \multicolumn{1}{l}{} & \multicolumn{1}{l|}{} & \multicolumn{1}{l}{} & \multicolumn{1}{l}{} \\ \hline
Daily (at least 1 time per day) &
   &
   &
  161 &
  40.4\% &
  123 &
  30.3\% \\ \hline
Weekly (1-5 times per week) &
   &
   &
  151 &
  37.8\% &
  152 &
  37.4\% \\ \hline
Monthly (1-5 times per month) &
   &
   &
  50 &
  12.5\% &
  58 &
  14.3\% \\ \hline
Less than once every month &
   &
   &
  23 &
  5.8\% &
  45 &
  11.1\% \\ \hline
Never &
   &
   &
  14 &
  3.5\% &
  28 &
  6.9\% \\ \hline
\multicolumn{1}{l}{\textbf{LLM Usage Tasks}}     &  &  & \multicolumn{1}{l}{} & \multicolumn{1}{l|}{} & \multicolumn{1}{l}{} & \multicolumn{1}{l}{} \\ \hline
Help with your academic writing (e.g., essays, research papers) &
   &
   &
  214 &
  53.6\% &
  212 &
  52.2\% \\ \hline
Help with your non-academic writing (e.g., emails, social media posts) &
   &
   &
  194 &
  48.6\% &
  163 &
  40.1\% \\ \hline
Help write code (e.g., debugging, generating code snippets) &
   &
   &
  91 &
  22.8\% &
  75 &
  18.5\% \\ \hline
Design or writing inspiration (e.g., brainstorming ideas, overcoming writer's block) &
   &
   &
  176 &
  44.1\% &
  184 &
  45.3\% \\ \hline
Independent research (e.g., exploring, building AI tools) &
   &
   &
  180 &
  45.1\% &
  166 &
  40.9\% \\ \hline
Translation (e.g., converting text between languages) &
   &
   &
  131 &
  32.8\% &
  136 &
  33.5\% \\ \hline
Search (e.g., finding information, looking up facts) &
   &
   &
  266 &
  66.7\% &
  257 &
  63.3\% \\ \hline
Mental health counseling (e.g., coping strategies, stress relief) &
   &
   &
  74 &
  18.5\% &
  88 &
  21.7\% \\ \hline
Entertainment/fun (e.g., casual conversations, creative storytelling) &
   &
   &
  158 &
  39.6\% &
  170 &
  41.9\% \\ \hline
Other &
   &
   &
  4 &
  1.0\% &
  13 &
  3.2\% \\ \hline
\multicolumn{1}{l}{\textbf{LLM Understanding}}   &  &  & \multicolumn{1}{l}{} & \multicolumn{1}{l|}{} & \multicolumn{1}{l}{} & \multicolumn{1}{l}{} \\ \hline
I have a substantive understanding of large language models. &
   &
   &
  130 &
  32.6\% &
  96 &
  23.6\% \\ \hline
I have read up on large language models and have a general understanding. &
   &
   &
  180 &
  45.1\% &
  181 &
  44.6\% \\ \hline
I have a vague understanding of how large language models work. &
   &
   &
  73 &
  18.3\% &
  107 &
  26.4\% \\ \hline
I don’t know much about how large language models work. &
   &
   &
  16 &
  4.0\% &
  22 &
  5.4\% \\ \hline
\end{tabular}%
}
\caption{Participant Demographic Table}
\Description{Table displaying demographic information for participants in two experimental setups: AAE LLM and QSLM. Includes identity, age, education, LLM usage frequency and tasks, and level of understanding of large language models. Values are presented as counts and percentages.}
\label{tab:demographicstable}
\end{table}

\section{Participant Demographic Table (includes those who failed manipulation check)}\label{ssec:demotablefail}
\begin{table}[H]
\centering
\resizebox{\columnwidth}{!}{%
\begin{tabular}{rllcc|cc}
\hline
\multicolumn{1}{l}{} &
   &
   &
  \multicolumn{2}{c|}{AAELM Setup (n=498)} &
  \multicolumn{2}{c}{QSLM Setup (n=487)} \\ \hline
\multicolumn{1}{l}{\textbf{Identity}}                   &  &  &     &        &     &        \\ \hline
Man                                                     &  &  & 268 & 41.6\% & 213 & 43.7\% \\ \hline
Woman                                                   &  &  & 224 & 37.3\% & 224 & 46.0\% \\ \hline
Non-binary                                              &  &  & 3   & 0.6\%  & 37  & 7.6\%  \\ \hline
Other                                                   &  &  & 3   & 0.6\%  & 13  & 2.7\%  \\ \hline
\multicolumn{1}{l}{\textbf{Age Group}}                  &  &  &     &        &     &        \\ \hline
18-24                                                   &  &  & 160 & 32.1\% & 108 & 22.2\% \\ \hline
25-34                                                   &  &  & 152 & 30.5\% & 205 & 42.1\% \\ \hline
35-44                                                   &  &  & 100 & 20.1\% & 83  & 17.0\% \\ \hline
45-54                                                   &  &  & 54  & 10.8\% & 55  & 11.3\% \\ \hline
55-64                                                   &  &  & 21  & 4.2\%  & 27  & 5.5\%  \\ \hline
65+                                                     &  &  & 11  & 2.2\%  & 9   & 1.8\%  \\ \hline
\multicolumn{1}{l}{\textbf{Education}}                  &  &  &     &        &     &        \\ \hline
Some high school                                        &  &  & 5   & 1.0\%  & 5   & 1.0\%  \\ \hline
High school or GED                                      &  &  & 74  & 14.9\% & 89  & 18.3\% \\ \hline
Associate                                               &  &  & 48  & 9.6\%  & 45  & 9.2\%  \\ \hline
Bachelor                                                &  &  & 251 & 50.4\% & 227 & 46.6\% \\ \hline
Master                                                  &  &  & 104 & 20.9\% & 102 & 20.9\% \\ \hline
PhD/MD/JD                                               &  &  & 11  & 2.2\%  & 15  & 3.1\%  \\ \hline
Other                                                   &  &  & 5   & 1.0\%  & 4   & 0.8\%  \\ \hline
\multicolumn{1}{l}{\textbf{LLM Usage Frequency}}        &  &  &     &        &     &        \\ \hline
Daily (at least 1 time per day)                         &  &  & 202 & 40.6\% & 167 & 34.3\% \\ \hline
Weekly (1-5 times per week)                             &  &  & 193 & 38.8\% & 179 & 36.8\% \\ \hline
Monthly (1-5 times per month)                           &  &  & 58  & 11.6\% & 64  & 13.1\% \\ \hline
Less than once every month                              &  &  & 31  & 6.2\%  & 47  & 9.7\%  \\ \hline
Never                                                   &  &  & 14  & 2.8\%  & 30  & 6.2\%  \\ \hline
\multicolumn{1}{l}{\textbf{LLM Usage Tasks}}            &  &  &     &        &     &        \\ \hline
Help with your academic writing (e.g., essays, research papers) &
   &
   &
  275 &
  55.2\% &
  212 &
  43.5\% \\ \hline
Help with your non-academic writing (e.g., emails, social media posts) &
   &
   &
  243 &
  48.8\% &
  236 &
  48.5\% \\ \hline
Help write code (e.g., debugging, generating code snippets) &
   &
   &
  127 &
  25.5\% &
  98 &
  20.1\% \\ \hline
Design or writing inspiration (e.g., brainstorming ideas, overcoming writer's block) &
   &
   &
  224 &
  45.0\% &
  228 &
  46.8\% \\ \hline
Independent research (e.g., exploring, building AI tools) &
   &
   &
  233 &
  46.8\% &
  204 &
  41.9\% \\ \hline
Translation (e.g., converting text between languages)   &  &  & 175 & 35.1\% & 167 & 34.3\% \\ \hline
Search (e.g., finding information, looking up facts)    &  &  & 336 & 67.5\% & 312 & 64.1\% \\ \hline
Mental health counseling (e.g., coping strategies, stress relief) &
   &
   &
  95 &
  19.1\% &
  113 &
  23.2\% \\ \hline
Entertainment/fun (e.g., casual conversations, creative storytelling) &
   &
   &
  193 &
  38.8\% &
  209 &
  42.9\% \\ \hline
Other                                                   &  &  & 6   & 1.2\%  & 13  & 2.7\%  \\ \hline
\multicolumn{1}{l}{\textbf{LLM Understanding}}          &  &  &     &        &     &        \\ \hline
I have a substantive understanding of large language models. &
   &
   &
  167 &
  33.5\% &
  138 &
  28.3\% \\ \hline
I have read up on large language models and have a general understanding. &
   &
   &
  221 &
  44.4\% &
  208 &
  42.7\% \\ \hline
I have a vague understanding of how large language models work. &
   &
   &
  93 &
  18.7\% &
  115 &
  23.6\% \\ \hline
I don’t know much about how large language models work. &  &  & 17  & 3.4\%  & 26  & 5.3\%  \\ \hline
\end{tabular}%
}
\caption{Participant Demographic Table (includes those who failed manipulation check)}
\Description{Table displaying demographic information for participants in two experimental setups: AAE LLM and QSLM. Includes identity, age, education, LLM usage frequency and tasks, and level of understanding of large language models. Values are presented as counts and percentages.}
\label{tab:demographicstablefail}
\end{table}


\section{Result Tables}\label{ssec:resulttables}


\begin{table}[!h]
\begin{tabular}{|l|c|c|}
\hline
                                & AAE Participants & Queer Slang Participants \\ \hline
Satisfaction w/ SAELM           & 0.113**             & 0.15**                    \\ \hline
Frustration w/ SAELM            & 0.173**             & 0.10**                     \\ \hline
Trust w/ SAELM                  & 0.133**             & 0.14**                     \\ \hline
Social Presence w/ SAELM        & 0.19**             & 0.20**                     \\ \hline
Satisfaction w/ Sociolect LM    & 0.05             & 0.12**                     \\ \hline
Frustration w/ Sociolect LM     & 0.123**             & 0.01                     \\ \hline
Trust w/ Sociolect LM           & 0.03             & 0.15**                     \\ \hline
Social Presence w/ Sociolect LM & 0.13**             & 0.19 **                    \\ \hline
Overall Reliance                & 0.01             & -0.02                    \\ \hline
Reliance w/ SAELM               & 0.02             & -0.04                    \\ \hline
Reliance w/ Sociolect LM        & -0.01            & 0            \\ \hline           
\end{tabular}
\caption{IDAQ Score Correlation with User LLM Perception and Reliance. 
\newline \textit{Note:} ** indicates significance at $p < .05$}
\Description{Table showing correlation values between IDAQ scores and user perceptions or reliance on LLMs for AAE and Queer Slang participants. Significant correlations include satisfaction, frustration, trust, and social presence with standard English LLMs, and some with sociolect LLMs, especially for Queer Slang users. Asterisks indicate statistical significance at p < .05.}
\label{tab:idaqcorrelation}
\end{table}

The Individual Differences in Anthropomorphism Questionnaire (IDAQ) measures a user's general tendency to attribute human-like characteristics to non-human entities. In our study, we hypothesized that participants with higher IDAQ scores—indicating a greater propensity for anthropomorphization—would demonstrate higher levels of reliance on the agents and more favorable perceptions of them. Specifically, we expected IDAQ scores to positively correlate with both reliance and perception measures. However, our results did not support this hypothesis. We found little to no correlation between participants' IDAQ scores and their perceptions of either the SAE-speaking or AAE-speaking language models, as well as their reported reliance on the agents. Although the IDAQ measures a general tendency to anthropomorphize non-human entities, participants' perceptions and reliance on the language models may have been shaped more by situational factors (e.g., the perceived task relevance, performance quality, or linguistic familiarity of the agents) rather than by their stable anthropomorphic tendencies. Because the sociolect-based language models were still recognizable as automated systems and participants had limited social interaction context beyond simple question-answering, the opportunity for anthropomorphic projection may have been constrained. As a result, IDAQ scores were not a strong driver of participants’ perceptions or reliance behaviors in this specific experimental setup. Several dimensions showed statistically significant correlations (denoted by **, p < .05). Notably, social presence with the SAELM was positively correlated with IDAQ scores for both AAE participants (r = .19**) and Queer Slang participants (r = .20**), suggesting that participants who perceived the agent as more socially present also tended to attribute human likeness to the LLMs. Future work should investigate whether enhancing anthropomorphic cues in LLM interactions increases users’ tendency to attribute human-like qualities to these systems.

\begin{table}[!h]
\begin{tabular}{|l|c|c|}
\hline
                                         & AAE      & Queer Slang \\ \hline
mean\_RelianceSAE                        & 0.719799 & 0.712315    \\ \hline
mean\_RelianceAAVE / mean\_RelianceQueer & 0.684461 & 0.687438    \\ \hline
p                                        & 0.033382 & 0.085205    \\ \hline
d                                        & 0.121212 & 0.080391    \\ \hline
\end{tabular}
\caption{RQ1: T-test for reliance}
\Description{Table comparing reliance scores on standard and sociolect LLMs for AAE and Queer Slang participants. Includes mean reliance values, p-values, and effect sizes from t-tests. AAE participants show a statistically significant difference in reliance, while the Queer Slang group does not.}
\label{tab:rq1reliance}
\end{table}




\begin{table}[!h]
\begin{tabular}{|l|l|l|}
\hline
                & AAE                                                                  & Queer Slang                                                                  \\ \hline
Trust           & \begin{tabular}[c]{@{}l@{}}p-value = 0.000627\\ d-value = 0.188291\end{tabular}  & \begin{tabular}[c]{@{}l@{}}p-value = 0.476128\\ d-value = -0.030697\end{tabular}         \\ \hline
Social presence & \begin{tabular}[c]{@{}l@{}}p-value = 0.187942\\ d-value = -0.058857\end{tabular} & \begin{tabular}[c]{@{}l@{}}p-value = 4.902323e-09\\ d-value = -2.342795e-01\end{tabular} \\ \hline
Satisfaction    & \begin{tabular}[c]{@{}l@{}}p-value = 0.000594\\ d-value = 0.200465\end{tabular}  & \begin{tabular}[c]{@{}l@{}}p-value = 0.210682\\ d-value = 0.065225\end{tabular}          \\ \hline
Frustration     & \begin{tabular}[c]{@{}l@{}}p-value = 0.000167\\ d-value = 0.216221\end{tabular}  & \begin{tabular}[c]{@{}l@{}}p-value = 0.000002\\ d-value = 0.245639\end{tabular}          \\ \hline
\end{tabular}
\caption{RQ2: T-tests with respective perception variables}
\Description{Table showing p-values and Cohen’s d-values for t-tests comparing perception variables—trust, social presence, satisfaction, and frustration—between AAE and Queer Slang conditions. Significant results are observed in the AAE group for trust, satisfaction, and frustration, while significant differences in the Queer Slang group appear for social presence and frustration.}
\label{tab:rq2}
\end{table}

\begin{figure}[htbp]
    \centering
    \begin{subfigure}{0.35\textwidth} 
        \centering
        \includegraphics[width=\linewidth]{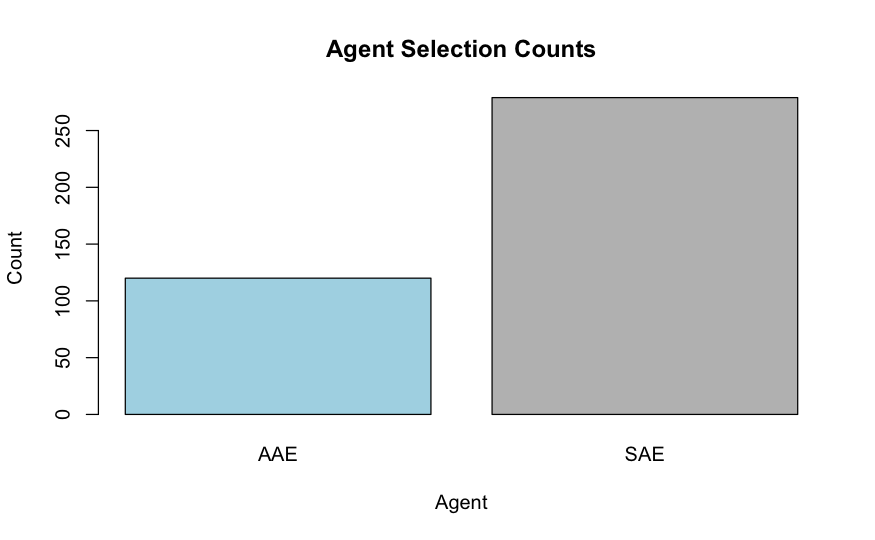}
        \caption{Preferences for AAE and SAE}
        \label{fig:aaebar}
    \end{subfigure}
    \hfill
    \begin{subfigure}{0.35\textwidth} 
        \centering
        \includegraphics[width=\linewidth]{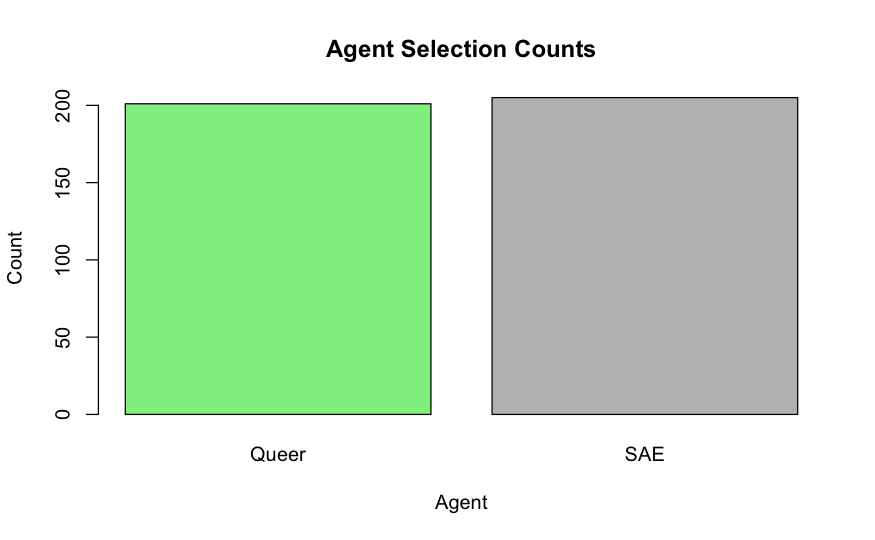}
        \caption{Preferences for Queer slang and SAE}
        \label{fig:queerbar}
    \end{subfigure}
    \caption {Participant preference for different language styles: (a) Bar graph showing participant preferences for SAE and AAE, with SAE being preferred significantly more. (b) Bar graph showing participant preferences for SAE and Queer slang, indicating nearly equal preference for both. Preference counts range from 0 to 300.}
    \Description{Two bar charts showing participant preferences for language models using either Standard American English or a sociolect. The left plot shows more participants selected the SAELM over the African American English  agent. The right plot shows near-equal preferences between the Queer slang and SAELMs.}
    \label{fig:combined_boxplots_2}
\end{figure}

\begin{table}[!h]
\begin{tabular}{|l|c|c|}
\hline
             & AAE          & Queer Slang \\ \hline
mean\_sample & 6.992481e-01 & 0.504926    \\ \hline
p-value           & 1.123979e-16 & 0.842924    \\ \hline
d-value            & 4.339398e-01 & 0.00984     \\ \hline
\end{tabular}
\caption{RQ2: T-test using Cohen's D 1-sample (with 0.5)}
\Description{Table showing results from a one-sample t-test using Cohen's D with a comparison value of 0.5. The AAE group shows a statistically significant difference with a p-value near zero and a moderate effect size. The Queer Slang group shows no significant difference.}
\label{tab:rq2preferredreliance}
\end{table}


\begin{table}[h!]
\centering
\begin{tabular}{|l|c|c|}
\hline
\textbf{Variable} & \textbf{AAE Reliance Correlation} & \textbf{SAE Reliance Correlation} \\
\hline
Trust & $r=0.256, \textbf{p<0.001}$ & $r=0.331, \textbf{p<0.001}$ \\ 
Social Presence & $r=0.188, \textbf{p<0.001}$ & $r=0.117, p=0.058$ \\ 
Satisfaction & $r=0.288, \textbf{p<0.001}$ & $r=0.377, \textbf{p<0.001}$ \\ 
Lack of Frustration & $r=0.195, \textbf{p<0.001}$ & $r=0.151, \textbf{p=0.008}$ \\ 
\hline
\end{tabular}
\caption{RQ3: Correlation between Language Reliance and Perception Variables for AAE Speakers. Significant results in bold.}
\Description{Table showing correlation coefficients between language reliance and perception variables—trust, social presence, satisfaction, and lack of frustration—for AAE and SAE speakers. Significant positive correlations are observed for most variables in both groups, with all AAE correlations significant and all but one SAE correlation also significant.}
\label{tab:rq3aaecorrelation}
\end{table}

\begin{table}[h!]
\centering
\begin{tabular}{|l|c|c|}
\hline
\textbf{Variable} & \textbf{QS Reliance Correlation} & \textbf{SAE Reliance Correlation} \\
\hline
Trust & $r=0.394, \textbf{p<0.001}$ & $r=0.376, \textbf{p<0.001}$ \\
Social Presence & $r=0.264, \textbf{p<0.001}$ & $r=0.133, \textbf{p=0.022}$ \\
Satisfaction & $r=0.438, \textbf{p<0.001}$ & $r=0.441, \textbf{p<0.001}$ \\
Lack of Frustration & $r=0.185, \textbf{p=0.001}$ & $r=0.070, p=0.265$ \\
\hline
\end{tabular}
\caption{RQ3: Correlation between Language Reliance and Perception Variables for Queer slang speakers. Significant results in bold.}
\Description{Correlation table comparing Queer Slang (QS) and Standard American English (SAE) reliance with four perception variables: trust, social presence, satisfaction, and lack of frustration. All correlations for QS are significant. For SAE, all except lack of frustration are significant.}
\label{tab:rq3queercorrelation}
\end{table}

\begin{table}[!h]
\begin{tabular}{|l|c|c|}
\hline
                           & AAE       & Queer Slang \\ \hline
mean\_RelianceDispreffered & 0.672431  & 0.66980     \\ \hline
mean\_ReliancePreferred    & 0.731830  & 0.731773    \\ \hline
p-value                         & 0.000325 & 0.000008   \\ \hline
d-value                          & -0.204427 & -0.207088   \\ \hline
\end{tabular}
\caption{RQ3: Table of Preference Reliance Results}
\Description{Table showing preference-based reliance results for AAE and Queer Slang conditions. Reliance scores are higher for preferred agents than dispreferred ones in both groups, with statistically significant differences.}
\label{tab:rq3explicitpreference}
\end{table}

\newpage

\section{AAE and Queer slang Codebook}\label{ssec:codebook}

{\small
\begin{longtable}
{|p{0.15\textwidth}|p{0.4\textwidth}|p{0.4\textwidth}|}
\hline
\multicolumn{3}{|c|}{SAELM} \\ \hline
\textbf{Code} & \textbf{Description} & \textbf{Example Comment} \\ \hline
Task-Oriented & Participants felt Agent Red stayed focused on the task without unnecessary distractions. & It seems more focused on the task at hand. Agent Blue sounds like a straight person attempting to write a gay person speaking. \\ \hline
Comprehensible & Participants feel they can understand Agent better & Easier to understand and not as influenced by vernacular \\ \hline
Bland & Participants felt Agent Red’s responses were bland, generic, or lacked personality, resembling corporate language. & Agent Red was boring and generic. I found myself smiling at Agent Blue's responses; they just felt warmer and more fun. \\ \hline
Formality & Participants preferred Agent Red’s formal and neutral tone, particularly for factual or professional interactions. & I expect a certain degree of formality when utilizing bots as opposed to more personal speech, so while the queer/AAVE slang doesn't bother me from the chatbot, it doesn't necessarily make me feel any more personal about it. I'd rather a robot didn't try to get too personable, especially when a lot of slang has roots in AAVE/queer culture and a chatbot can't fully understand the implications and history of what it's saying. It feels too forced, especially if I just need a factual answer. \\ \hline
\multicolumn{3}{|c|}{QSLM} \\ \hline
\makecell[l]{Expectation\\Misalignment} & Participants felt Agent Blue’s human-like traits conflicted with their expectations of AI behavior. & I don't want robots talking like humans \\ \hline
\makecell[l]{Anthropomorphi\\-zation} & Participants perceived Agent Blue as more human, personable, with a social presence. & Agent Blue comes across as more human and less formal. \\ \hline
Not Normal & Participants’ mental model of AI did not align with sociolect usage, leading to discomfort or skepticism. & When using AI, I do not need it to sound like a human being. \\ \hline
Resonates & Participants felt a personal connection to Agent Blue and appreciated its familiar and casual tone. & I enjoy being called diva! \\ \hline
Disrespectful & Participants felt Agent Blue’s use of slang (queer/AAVE) was inappropriate, stereotyping, or mocking the associated culture. & the blue seems to be making a mockery of lgtb \\ \hline
Exaggerated Usage & Participants felt the use of slang was excessive and did not reflect natural usage. & Even people who use LGBTQ slang don't talk like that constantly. It would be annoying to have an AI constantly use slang phrases. \\ \hline
Negative Emotions & Vocab participants have used to describe how they felt: annoying, laughable, irritating, enraging & Even people who use LGBTQ slang don't talk like that constantly. It would be annoying to have an AI constantly use slang phrases. \\ \hline
Positive Emotions & Vocab participants have used to describe how they felt: fun, pleasing & It's just more fun to hear/read \\ \hline
Context & Participants felt that Agent's usage of vernacular did not align with context/ topic of discussion & While queer slang is perfectly fine, like any slang there's a time and a place to use it. This situation doesn't feel like it. \\ \hline
Unfamiliar & Participants felt that Queer slang may not be understood by some users and serve as a potential barrier to Agent usage & This is because not everyone is able to understand these slang terms, and misinformation with someone who is unfamiliar with them could result. \\ \hline
\caption{Codebook generated during open coding of Queer slang participant comments}
\Description{Codebook table outlining qualitative themes from participant comments on AAE and Queer slang language models. Codes include perceptions of task orientation, anthropomorphism, excessive usage, and emotional responses, with example quotes for each.}
\label{tab:queercodes}
\end{longtable}

\begin{longtable}{|p{0.15\textwidth}|p{0.4\textwidth}|p{0.4\textwidth}|}
\hline
\multicolumn{3}{|c|}{SAELM} \\ \hline
\textbf{Code} & \textbf{Description} & \textbf{Example} \\ \hline
Normal & Agent Red aligns with participants' expectations of typical AI behavior. & Agent Blue using AAVE sounds like a joke and not natural. I feel like I wouldn't take the situation seriously using Blue because I'm not speaking with an actual Black person. Agent Red sounds normal and like an actual AI. \\ \hline
Comprehensible & Participants perceive Agent Red as clear, easy to understand, and capable of facilitating quality conversations. Provides an impression of being more educated and confident. & I would prefer to interact with Agent Red because I think the conversation would be more productive. \\ \hline
Formality & Participants preferred Agent Red’s formal and neutral tone, particularly for factual or professional interactions. & Because he seems more professional. \\ \hline
Resonates & Participants find Agent Red personable, relatable, and casual. & Both are okay, but I prefer Agent Red more because I have a personal connection with him. \\ \hline
Reliable & Participants felt that they could rely on responses of Agent Red versus those of Agent Blue. & Agent Red sounds more reliable and trustworthy. \\ \hline
Confident & Participants believe Agent Red exudes confidence in their response. & Because it sounded sure with its answers. \\ \hline
Enthusiastic & Participants find Agent Red more eager, proactive, enthusiastic, and motivated to do the task for them. & Agent Red because it expressed enthusiasm and willingness to help. \\ \hline
\multicolumn{3}{|c|}{AAELM} \\ \hline
\textbf{Code} & \textbf{Description} & \textbf{Example} \\ \hline
Unserious & Participants perceive Agent Blue as a prank or joke. & Agent Blue sounds like a prank. \\ \hline
Unnatural & Participants feel Agent Blue’s use of AAVE is forced or abnormal. & Agent Blue using AAVE sounds like a joke and not natural. I feel like I wouldn't take the situation seriously using Blue because I'm not speaking with an actual Black person. Agent Red sounds normal and like an actual AI. \\ \hline
Not normal & Agent Blue challenges participants' mental model of AI by incorporating sociolects. & Agent Blue using AAVE sounds like a joke and not natural. I feel like I wouldn't take the situation seriously using Blue because I'm not speaking with an actual Black person. Agent Red sounds normal and like an actual AI. \\ \hline
Positive Emotions & Vocab participants have used to describe how they felt: fun, pride. & It just sounds more fun to interact with. I would be able to speak with it casually. \\ \hline
Resonates & Participants find Agent Blue personable, relatable, and casual. & I prefer Agent Blue because I feel closer to it. This is the first time I am seeing an AI interact with an African American vernacular. It felt very relatable. \\ \hline
Disrespectful & Participants feel Agent Blue’s use of AAVE mocks or stereotypes Black language and culture. & I feel like when people try to use AAVE who aren't Black, it comes across as disrespectful. I don't like the idea of AI using such language, especially if it is being programmed by someone who isn't Black. \\ \hline
Negative Emotions & Vocab participants have used to describe how they felt: annoying, distrust. & I would never be able to use Agent Blue because I would be annoyed the entire time. Yes, there are some Black people who speak like that, but here it feels insulting. \\ \hline
Improper & Participants associate Agent Blue's language with an improper form of English. & Agent Red sounds like an intelligent being, whereas Agent Blue sounds like they don't know how to speak proper English. \\ \hline
\makecell[l]{Anthropomorphi\\-zation} & Participants perceived Agent Blue as more human, personable, with a social presence. & Her answers felt more like me talking. \\ \hline
Better Sounding & Participants perceive Agent Blue just sounding better. & It's straightforward, the less formal sounds better. \\ \hline
Context & Participants felt that Agent's usage of vernacular did not align with context/topic of discussion. & Agent Red sounds more professional and coherent, so I would prefer Agent Red in a more professional setting. For more casual interactions, either agent would do. \\ \hline
Comprehensible & Participants perceive Agent Blue as clear, easy to understand, and capable of facilitating quality conversations. Provides an impression of being more educated and confident. & Agent Blue is brief and therefore concise and precise. \\ \hline
Reliable & Participants felt that they could rely on responses of Agent Blue versus those of Agent Red. & Because Agent Blue was honest, reliable, etc. \\ \hline
Confident & Participants believe Agent Blue exudes confidence in their response. & Agent Blue was sure about its opinion. \\ \hline
Enthusiastic & Participants find Agent Blue more eager, proactive, enthusiastic, and motivated to do the task for them. & This response is polite, professional, and clear. It shows willingness to help and provides a helpful and informative answer. \\ \hline
\caption{Codebook generated during open coding of AAE participant comments}
\Description{Codebook table outlining qualitative themes from participant comments on AAE and Queer slang language models. Codes include perceptions of task orientation, anthropomorphism, excessive usage, and emotional responses, with example quotes for each.}
\end{longtable}
}

\newpage
\section{Query Pseudocode Generation}\label{ssec:codecount}

\begin{table*}[h!]
    \centering
    \begin{minipage}[t]{0.48\textwidth}
        \centering
        \begin{tabular}{l r r}
            \toprule
            \textbf{Code} & \textbf{Value} & \textbf{Percentage}\\
            \midrule
            Positive Emotions & 156& 32.0\%\\
            Resonates & 45& 9.2\%\\
            Negative Emotions & 40& 8.2\%\\
            Anthropomorphization & 38& 7.8\%\\
            Exaggerated Usage & 34& 7.0\%\\
            Disrespectful & 26& 5.3\%\\
            Not normal & 18& 3.7\%\\
            Unfamiliar & 17& 2.9\%\\
            Context & 14& 2.0\%\\
            Expectation Misalignment & 14& 2.9\%\\
            \bottomrule
        \end{tabular}
        \caption{Codes generated pertaining to QSLM while coding participant feedback for the QSLM Setup.}
        \Description{Table showing the frequency and percentage of qualitative codes derived from participant feedback about the QSLM setup. The most common codes include Positive Emotions, Resonates, and Negative Emotions.}
        \label{tab:agentbluequeer}
    \end{minipage}
    \hfill
    \begin{minipage}[t]{0.48\textwidth}
        \centering
        \begin{tabular}{l r r}
            \toprule
            \textbf{Code} & \textbf{Value} & \textbf{Percentage}\\
            \midrule
            Formality & 60 & 12.3\%\\
            Comprehensible & 36 &7.4\%\\
            Bland & 28 & 5.8\%\\
            Task-Oriented & 24 & 4.9\%\\
            \bottomrule
        \end{tabular}
        \caption{Codes generated pertaining to SAELM while coding participant feedback for the QSLM Setup.}
        \Description{Table showing qualitative codes from participant feedback about the SAELM setup. Most common codes include Formality, Comprehensible, Bland, and Task-Oriented, with Formality being the most frequent.}
        \label{tab:agentredqueer}
    \end{minipage}
\end{table*}

\begin{table*}[h!]
    \centering
        \begin{minipage}[t]{0.48\textwidth}

    \begin{tabular}{l r r}
        \toprule
        \textbf{Code} & \textbf{Value}  & \textbf{Percentage} \\
        \midrule
        Resonates & 44  & 8.8\%\\
        Anthropomorphization & 22& 4.4\%\\
        Positive Emotions & 21 & 4.2\%\\
        Unnatural & 17 & 3.4\%\\
        Comprehensible & 13 & 2.6\%\\
        Unserious & 13 & 2.6\%\\
        Reliable & 10& 2.0\%\\
        Improper & 8 & 1.6\%\\
        Disrespectful & 7& 1.4\% \\
        Negative Emotions & 4 & 0.8\%\\
        Better Sounding & 3& 0.6\% \\
        Context & 3& 0.6\%  \\
        Not normal & 2& 0.4\%\\
        Formality & 1& 0.2\% \\
        Confident & 1& 0.2\% \\
        \bottomrule
    \end{tabular}
    \caption{Codes generated pertaining to AAELM while coding participant feedback for the AAELM Setup.}
    \Description{Table showing qualitative codes from participant feedback about the AAELM setup. Top themes include 'Resonates', 'Anthropomorphization', and 'Positive Emotions', with less frequent mentions of traits like 'Confident' and 'Formality'.}
    \label{tab:agentblueaae}
\end{minipage}
    \hfill
        \begin{minipage}[t]{0.48\textwidth}
    \centering
    \begin{tabular}{l r r}
        \toprule
        \textbf{Code} & \textbf{Value} & \textbf{Percentage}\\
        \midrule
        Comprehensible & 78 & 15.6\%\\
        Formality & 58 & 11.6\%\\
        Reliable & 27& 5.4\%\\
        Normal & 19 & 3.8\%\\
        Confident & 9& 1.8\%\\
        Resonates & 7 & 1.4\%\\
        Anthropomorphization & 6 & 1.2\%\\
        Enthusiastic & 6& 1.2\%\\
        Task-Oriented & 4 &0.8\%\\
        \bottomrule
    \end{tabular}
    \caption{Codes generated pertaining to SAELM while coding participant feedback for the AAELM Setup.}
    \Description{Table showing participant feedback codes for the SAELM setup in the AAELM condition. Most frequent codes include 'Comprehensible', 'Formality', and 'Reliable', while 'Enthusiastic' and 'Task-Oriented' were among the least mentioned.}
    \label{tab:agentredaae}
    \end{minipage}
\end{table*}









\end{document}